\documentclass[useAMS]{mn2e_modmargin}
\usepackage{amsmath}
\usepackage{url}
\usepackage{amsfonts}
\usepackage{amsbsy}
\usepackage{subfigure}
\usepackage{verbatim}
\usepackage{amssymb}
\usepackage{amsmath}
\usepackage{amsbsy}
\usepackage{graphicx}
\usepackage{array}
\usepackage[usenames, dvipsnames]{color}

\title{Dissipative structures in magnetorotational turbulence}
\author[J. Ross \& H.N. Latter]{Johnathan Ross$^{1}$,
   Henrik N. Latter$^{1}$\thanks{E-mail:
   hl278@cam.ac.uk}\\
$^{1}$ DAMTP, University of Cambridge, CMS, Wilberforce Road
Cambridge CB3 0WA, UK\\
}
\date{}

\begin{document}

\maketitle

\begin{abstract}

Via the process of accretion, magnetorotational turbulence
removes energy from a disk's orbital motion and transforms it into heat. 
Turbulent heating is far from uniform and is usually concentrated
in small regions of intense dissipation, characterised by abrupt
magnetic reconnection and higher temperatures. These regions are of
interest because they might generate non-thermal emission, in the form of
flares and energetic particles, or thermally process solids
in protoplanetary disks. Moreover, the nature of the
dissipation bears on the fundamental dynamics of the magnetorotational
instability (MRI) itself: local simulations indicate that the
large-scale properties of the turbulence
(e.g.\ saturation
levels, the stress-pressure relationship) depend
 on the short dissipative scales. In this paper
we undertake a numerical study
of how the MRI dissipates and the
small-scale dissipative structures it employs to do so. We use the
Godunov code RAMSES and unstratified compressible shearing boxes.
Our simulations reveal that dissipation is concentrated in ribbons of strong magnetic
reconnection that are significantly elongated in azimuth, up to a scale
height. Dissipative structures are hence meso-scale objects, and
potentially provide a route by which large scales and
small scales interact. We go on to show how these ribbons
evolve over time --- forming, merging, breaking apart, and
disappearing. 
Finally,
we reveal important couplings between
the large-scale density waves generated by the MRI and the
small-scale structures, which may
illuminate the stress-pressure relationship in MRI turbulence. 

\end{abstract}


\section{Introduction}

The turbulent angular momentum transport observed in sufficiently hot or ionised
accretion disks is most plausibly supplied by the magnetorotational
instability (MRI; Balbus and Hawley 1991, 1998). 
By permitting material in the disk to accrete, MRI
turbulence drives the intense luminosity of numerous astrophysical
sources, AGN most
notably, while regulating the growth and evolution of the central
object. Meanwhile, it influences several
separate processes, such as planet formation, outflows, outbursts,
warping, and magnetic flux transport (e.g.\ Nelson and Gressel 2010,
Fromang et al.~2013, Gammie 1996, Latter and Papaloizou 2012, Ogilvie
1999, Guilet and Ogilvie 2012). 

In the process of transporting angular
momentum and mass, the MRI's turbulent stress extracts energy
from the orbital shear which, after tumbling down a
turbulent cascade, degrades into heat by Ohmic and viscous
dissipation. 
At least in thin disks, the preponderance of
this energy is radiated away, to be intercepted by
astronomical
instruments. For many purposes, the
details of the thermalisation of orbital energy is
unimportant and an alpha viscosity model suffices to reproduce the
disk's broad-brush features (accretion, structure, spectra, etc). 
However, there are applications and contexts for which we might want to 
probe a little deeper.

To begin,
turbulent heating is rarely uniform, with intense 
dissipation (reconnection, in particular)
 taking place in localised regions (e.g.\ sheets and vortices)
and almost no dissipation in the large volumes between (e.g. Uritsky
et al.~2010). Rapid reconnection in these regions
could produce unexpectedly energetic and non-thermal emission, which might
relate to observed variability, flares, and even particle
acceleration (McClintock and Remillard 2006,
Belloni 2010, Yuan and Narayan 2014). The low filling factor of dissipation
also means narrow filaments of gas can be significantly hotter than 
their surroundings. Thus spatially and temporally intermittent dissipation
in protoplanetary disks might provide a way to
thermally process chondrule-precursors in gas that is too cool on the
average to do so (e.g.\ McNally et al.~2013).

More generally, the issue of
dissipation is important in establishing how the MRI operates.
Recent local simulations show that large-scale features of the turbulent
flow depend on the short dissipative scales, especially in the absence of
net-flux magnetic fields. 
Saturation levels, the magnitude of alpha, and the pressure-stress
relationship all appear to depend on the size of the magnetic Prandtl number and
whether explicit diffusion is used or not
(Fromang and Papaloizou 2007, Fromang et al. 2007, Simon
et al.~2009, Riols et al.~2013, Meheut
et al.~2015, Ross et al.~2016, Ryan et al.~2017). It is yet unclear if
this non-locality in wavenumber space carries over to realistic 
astrophysical flows, 
where the separation of scales is vast compared
with what is possible in simulations (see discussion in Lesur and
Logaretti 2011). It is hence essential to understand \emph{physically}
the interaction
between the small and large scales in simulations so as to help us 
understand this issue better. 

We undertake local box simulations of the MRI using the finite-volume 
Godonov code RAMSES (without its AMR capabilities) 
(Fromang et al.~2006). 
Our focus is the characterisation and evolution of
dissipative structures, and so we describe their geometry and
heating intensity at given instants (snapshots), and how these
properties vary over time. 
We also make a start exploring their relationship with
compressibility. 
To perform the
analysis we adopt the techniques and diagnostics
developed by Uritsky et al.~(2010), Zhdankin et al.~(2013),
Momferratos et al.~(2014), and Zhdankin et al.~(2015a, 2015b). 
Recently, Zhdankin et al.~(2017; hereafter ZWBL17) presented similar
results but for an incompressible gas and mean magnetic field; they
also restricted themselves to a `spatial analysis', i.e. to snapshots.
We generalise these results to compressible flow (thus
bringing in an explicit outer scale, the disk thickness $H$) and
zero-net-flux fields. We also track the evolution of dissipative
structures over time. 

Our simulations show that these features exhibit a 
characteristic geometry akin to thin ribbons,
elongated in azimuth, but canted at some angle $\sim
10^\circ$. While the thinness and width of the
ribbons are small
and controlled by the diffusivities, their length can be remarkably
large, with an average $\lesssim H$.
In our simulations, dissipative structures are hence
meso-scale, and not necessarily tiny nor neglectable. 
In particular, their elongation may present an avenue whereby
small-scale dissipation influences the large-scales. Moreover, intense heating
events associated with the biggest structures may have observable
consequences, possibly exciting 
low-level variability in luminosity.
Separately, we construct statistics and scaling laws for the 
heating rate, which accord with those derived from 
reduced MHD (RMHD), in agreement with ZWBL17;
though the majority of our simulations are zero-net-flux simulations, 
strong azimuthal fields do develop that can function locally
as a guide field for the small-scales.

The hot dissipating ribbons of gas are dynamic in time ---
forming, merging, breaking apart, and evaporating. The 
longest-lived structures tend to be the biggest, and during an orbit can
dissipate considerable energy, potentially causing significant
temperature inhomogeneities. The largest structures start breaking up
midway through their lives, possibly through the action of
instabilities of Kelvin-Helmholtz or tearing type (Loureiro et
al.~2007, 2013). We also uncover intriguing couplings
between the dissipating ribbons and both small-scale acoustic waves and
large-scale density waves. Of special interest is how the passage of a
density wave through a structure intensifies heating within it,
while concurrently distorting the wave. This is
a clear example of how small-scale dissipation and
large-scales can interact.

The paper is organised as follows. The model equations, numerical
approach, and setup are presented in Section 2. Significant space
is devoted to the diagnostic tools developed by Zhdankin et
al.~(2013, 2015b) which we use to identify dissipative
structures, characterise their spatial
properties, and track them over
time. Sections 3, 4, and 5 present our results, which are split into a
spatial analysis of structures at given moments, a temporal
analysis which describes their evolution, and a brief exploration of
their relationship with shocks and density waves. We draw our conclusions in
Section 6.

\section{Governing equations, numerical tools, 
    and diagnostics}

\subsection{Formulation}

We wish to explore the essential small-scale
features of the turbulent flow
and so we adopt an idealised local set-up, 
the shearing box model (Goldreich \&
Lynden-Bell 1965, Latter \& Papaloizou 2017). 
It describes a small Cartesian chunk of disk embedded near
the midplane where locally the differential rotation appears as a
linear shear flow plus rotation, and the vertical gravity of the
central object can be omitted.
The angular frequency vector of the corotating frame is
$\mathbf{\Omega}=\Omega\boldsymbol{\hat{e}_{z}}$.
As is customary, $x,y,z$
are the radial, azimuthal, and vertical spatial variables in the
shearing box, and
$\boldsymbol{\hat{e}}_{x}$, $\boldsymbol{\hat{e}}_{y}$,
$\boldsymbol{\hat{e}}_{z}$ are the corresponding unit vectors. 
The gas located in the shearing sheet is governed by
the equations of compressible MHD:
\begin{align}
\label{eq::1}
&\frac{\partial \rho}{ \partial t} + \nabla \cdot ( \rho \boldsymbol{v}
) = 0, \\
&\rho \frac{\partial \boldsymbol{v}}{\partial t} +\rho (\boldsymbol{v}
\cdot \nabla)\boldsymbol{v} 
 = - 2 \rho \bold{\Omega} \times \boldsymbol{v}  +
3x\rho \Omega^{2}\boldsymbol{\hat{e}}_{x}-\nabla P \notag\\
& \hskip3cm  +( \nabla \times \boldsymbol{B} ) \times \boldsymbol{B}+\nabla\cdot\boldsymbol{\Pi} \\
&\frac{\partial \boldsymbol{B}}{\partial t} = \nabla \times
(\boldsymbol{v} \times \boldsymbol{B})+\eta\nabla^2\boldsymbol{B}, 
\end{align}
where $\rho$ is the mass density, $\boldsymbol{v}$ is the velocity, 
$P$ is the gas pressure, and $\boldsymbol{B}$ is the magnetic
field. The (molecular) viscous stress is given by
$$ \boldsymbol{\Pi}= \rho\nu\left[\nabla\boldsymbol{v}+
(\nabla\boldsymbol{v})^\text{T}-\frac{2}{3}(\nabla\cdot\boldsymbol{v})\mathbf{1}\right], $$
and $\nu$ is the constant kinematic viscosity. The constant magnetic
diffusivity is $\eta$.

For most of the simulations in this paper these equations are 
closed by the \emph{isothermal} equation of
state $P=c_s^2\rho$, with $c_s$ the constant isothermal speed of
sound. However, the small set of simulations shown in Section 5
 are diabatic with the internal
energy $\varepsilon$ determined from
\begin{equation}
\frac{\partial \varepsilon}{\partial t}+\boldsymbol{v}\cdot
\nabla\varepsilon = -P\nabla\cdot \boldsymbol{v} + Q - \Lambda,
\end{equation}
where viscous and Ohmic heating is represented by $Q$ and cooling
by $\Lambda$, a simple relaxation law $\Lambda=\theta
(P-P^*)/[(\gamma-1)\tau_c]$,
where, $\gamma$ is the adiabatic index,
$\tau_c$ is a relaxation timescale, and $P^*$ is the pressure the
system wants to cool to, chosen so that the quasi-steady turbulent
state settles on a temperature approximately equal to the initial
temperature. The relaxation time is set to $\tau_c=5/\Omega$. 
In this case we adopt an ideal gas
equation of state, and so $\varepsilon= P/(\gamma-1)$.

\subsection{Numerical methods and set-up}

All of the simulations that we perform are carried out using RAMSES, 
a finite-volume Godunov code based on the MUSCL-Hancock algorithm 
(Teyssier 2002; Fromang et al.~2006), but with its AMR capabilities
disabled (i.e. the `Dumses' version).
We employ the HLLD Riemann solver 
(Miyoshi \& Kusano 2005),
 and the multidimensional slope limiter described in Suresh (2000).
 For further
 details of our numerical scheme see Ross et al.~(2016).

For the all the simulations shown, we used a box size of
$(1,5,1)H$  with a resolution vector of
$N=(N_{x},N_{y},N_{z})= (1,2,1)n$ where $n \in
\lbrace 64,128, 256\rbrace$. The scale height of the disk is
$H=c_s/\Omega$. 
The grid lengths are denoted by
$\delta_x$, $\delta_y$, and $\delta_z$.  

Three initial configurations of magnetic field were trialled: (a) zero
net-flux, for which $\boldsymbol{B}=B_{0} \sin (2 \pi 
x)\boldsymbol{\hat{e}}_{z} $,
(b) net-toroidal flux,
 $\boldsymbol{B}=B_{0}\boldsymbol{\hat{e}}_{y}$,
and (c)
 net-vertical flux, $\boldsymbol{B}=B_{0}\boldsymbol{\hat{e}}_{z}$.
To induce the MRI we introduce random velocity perturbations in all
principle directions with amplitudes $ < 0.1c_{s}$.
We choose code units so
that $c_s=10^{-3}$ and $\Omega=10^{-3}$. 
Thus the plasma beta in code units is
$\beta=2/B_0^2$.

The relative sizes of viscous and Ohmic diffusion are quantified
by the Reynolds and magnetic Reynolds numbers: $\text{Re}= H^2\Omega/\nu$ and
$\text{Rm}=H^2\Omega/\eta$. The magnetic Prandlt number is defined to be
their ratio: Pm= Rm/Re. In previous work a resolution of 128$/H$ is
deemed adequate for Reynolds numbers of roughly 6000, and a resolution
of 256$/H$ for numbers twice that (e.g.\ Fromang et
al.~2007). However, it is likely that even better resolution is
necessary to accurately describe small-scale
dissipative phenomena (as opposed to quantities like $\alpha$). 
We note that ZWBL17 employ $N_z=512$ grid zones to
represent simulations with $\text{Rm}=45000$. However, to completely mitigate 
grid pollution, we employ much lower Reynolds numbers,
taking Rm=5000 and Pm=4. (Another consideration is
that finite volume codes are less accurate than spectral codes for a
given resolution.)

The results of five simulations are presented. Three
are zero-net-flux simulations with resolutions $n=64$, 128, and 256, labelled
`znf64', `znf128', and `znf256'. A single vertical-net-flux simulations is shown 
with $n=128$ and $\beta=200$, labelled `vert128', and a toroidal-net-flux simulation
with $n=128$ and $\beta=400$, labelled `tor128'.

\subsection{Spatial characterisation of dissipative structures}

We outline the main diagnostics to
(a) identify coherent dissipative structures in the simulations, and (b)
 measure their fundamental properties. In this subsection we deal
with the extraction and analysis of these structures at fixed times,
i.e. at snapshots. In the subsequent subsection we
describe how to analyse their temporal evolution. The method follows that of
Zhdankin et al.~(2013, 2015b).

\subsubsection{Identification}

The single most important part of the analysis is to distinguish, within
the disordered turbulent flow, coherent spatially connected
regions of the fluid that dissipate at a level far greater than the
average. These special blobs of fluid we identify as dissipative
structures, the regions in which most of the turbulent energy is thermalised. 

To begin, we quantify the 
local resistive and viscous dissipation
 by $\epsilon_{\eta}(\boldsymbol{x})=\eta|\boldsymbol{J}|^2=\eta|\nabla
\times\mathbf{B}|^{2}$ and
$\epsilon_{\nu}(\boldsymbol{x})=\boldsymbol{\Pi}:\nabla\boldsymbol{v}$. 
In a given snapshot it is possible to compute the
spatial means of these dissipation rates, $\mu_\eta$ and $\mu_\nu$,
respectively, in addition to their standard deviations $\sigma_\eta$
and $\sigma_\nu$. We next construct the quantities 
\begin{align}
\epsilon_{\eta}^{k}= \mu_{\eta}+k\sigma_{\eta}, \qquad
\epsilon_{\nu}^{k} = \mu_{\nu}+k\sigma_{\nu},
\end{align}
which are the Ohmic and viscous dissipation rates greater than their
respective means by $k$-sigma, where $k$ is a free parameter we can
choose. These quantities serve as the boundaries above
which we deem a dissipative structure to be active. Formally, a region, or
structure, of high dissipation is then defined to be a spatially 
connected set of points $\boldsymbol{x}$ that satisfy one or both of the conditions
\begin{equation}\label{Eqn::etaCondition}
\epsilon_\eta(\boldsymbol{x}) > \epsilon_\eta^k, \qquad \epsilon_\nu(\boldsymbol{x}) >
\epsilon_\nu^k. 
\end{equation}
In practice we only use the first condition, as Ohmic
dissipation in MRI turbulence is so dominant. 

Within our simulation code, we have written an algorithm that at any
given instant, computes both $\epsilon_\eta(\boldsymbol{x})$ and
$\epsilon_\eta^k$ and by applying \eqref{Eqn::etaCondition} thus identifies
all the grid points of high dissipation. Coherent spatially adjoining
points we group together and identify as a dissipative structure.
To distinguish the coherent structures, we index them with an
integer, $i$, and represent them by $\Lambda_{i}$, a spatially  
connected set of points (grid cells). Note that the 
 algorithm takes into account the periodicity of the domain: a
structure on the boundary is matched with a structure at the
spatially identified points on the opposite boundary.

Lastly, we
impose
an additional filter to screen out tiny short-lived structures the size of a
few grid-cells. These we regard as artifacts of the grid, and probably
due to insufficient resolution. Structures of a width below some
threshold $\sim 4 \delta_x$, are discarded from the analysis. See below
for a quantitative definition of a structure's `width'.

\subsubsection{Structure analysis}

Once the structures have been identified, various measurements can be
made that outline their geometry and other properties.

First, we define the primary `length' of a structure, $L$, the
maximum distance between two points in $\Lambda_i$:
\begin{align}
L_{i}=\max_{\mathbf{p},\mathbf{q}\in\Lambda_{i}}\left\{|\mathbf{p}-\mathbf{q}|\right\},
\end{align}
with associated vectors $\mathbf{L}_{i}$ and
$\mathbf{\hat{L}}_{i}=\mathbf{L}_{i}/L_{i}$.
Here, the maximum is with respect to the Eulerian distance. The
subscript $i$ refers to the $i$th structure. Unless
otherwise stated, $L$ will refer to the length of the structure and
\textit{not} the box size.

Second, we measure the `width' of the
structure, defined as the longest distance between two cells in $\Lambda_i$
whose separation vector is perpendicular to $\hat{\mathbf{L}}_{i}$:
\begin{align}
W_{i}=\max_{\mathbf{p},\mathbf{q}\in\Lambda_{i}}\left\{|\mathbf{p}-\mathbf{q}|\,:\, \hat{\mathbf{L}}_{i}\cdot
(\mathbf{p}-\mathbf{q})=0\right\} 
\end{align}
with associated vectors $\mathbf{W}_{i}$ and
$\mathbf{\hat{W}}_{i}=\mathbf{W}_{i}/W_{i},$

Finally, the longest direction perpendicular to both the width and length can
be defined through:
\begin{align}
T_{i}=\max_{\mathbf{p},\mathbf{q}\in\Lambda_{i}}\left\{|\mathbf{p}-\mathbf{q}| \,:\, \mathbf{\hat{L}}_{i}\cdot
(\mathbf{p}-\mathbf{q})=0 \cap \mathbf{\hat{W}}_{i}\cdot
(\mathbf{p}-\mathbf{q})=0\right\},
\end{align} 
with associated vectors $\mathbf{T}_{i}$
and $\mathbf{\hat{T}}_{i}=\mathbf{T}_{i}/T_{i}$. This is referred to
as the `thickness'. 

In addition to these quantities, 
we measure the volume $V_i$, area $A_i$, and power $D_i$ dissipated by
each structure $\Lambda_i$:
\begin{align}
V_{i}&=\delta_{x}\delta_{y}
\delta_{z}\sum_{\mathbf{p}\in\Lambda_{i}}1, \\
A_{i}&= \sum_{\mathbf{p}\in\Lambda_{i}}X_{\mathbf{p}}\delta_{y}\delta_{z} 
+\sum_{\mathbf{p}\in\Lambda_{i}}Y_{\mathbf{p}}\delta_{x}\delta_{z} +
\sum_{\mathbf{p}\in\Lambda_{i}}Z_{\mathbf{p}}\delta_{x}\delta_{y} \\
D_{i}&=\delta_{x}\delta_{y}\delta_{z}
\sum_{\mathbf{p}\in\Lambda_{i}}\epsilon(\mathbf{p}),
\end{align}
where $X_{\mathbf{p}}$, $Y_{\mathbf{p}}$ and $Z_{\mathbf{p}}$ are the 
number of exterior faces of cell $\mathbf{p}$ and recall that
 $\delta_{x},\delta_{y},\delta_{z}$ are the grid lengths.

\subsection{Temporal characterisation
    of dissipative structures} 

We now present diagnostics and ideas to help understand how the
structures identified in the previous subsection evolve over time
and interact. This requires not only
the identification of a structure at a given time, but
also the identification of the same structure at adjacent times, as it
moves and changes. 
Note that finite data storage restricts the number of outputs (snapshots),
putting limitations on any temporal analysis. Only moderate cadences
and durations can be analysed for high resolution
simulations. 

At any instance in time, once structures are extracted
and measured, they must be related to the structures identified in
the previous snapshot, 
through a mapping which need not be bijective (due to
merging and dividing, for example). Therefore, the cadence
must be sufficiently high for this process to be possible. However,
this can only be an approximation and grid-sized structures are
problematic
 to track. Larger and longer lived features, however, can be
 characterised adequately. The time increment between outputs is
 referred to as $\Delta t$.

\subsubsection{Terminology and procedure}

Following Zhdankin et al.~(2015b), we use the following terminology.
A `state' is an individual spatial structure at a fixed time. A `path' 
 is a sequence of states over time, describing the evolution of what
 we regard as the same coherent structure with no merging or division.
The states within a path hence may be connected by a bijective
mapping, except at the endpoints of the path.
 A `process' is a collection of paths
 that are connected at their endpoints, and thus corresponds to a
 group of dissipative structures that interact over time. Finally,
a `complete path' is one whose existence (formation and destruction)
is contained within the duration of our temporal analysis.

Suppose we have $N_\text{out}$ snapshots of our turbulent flow, each taken
$\Delta t$ apart in time. The basic method to identify paths and processes
is as follows. In each snapshot we conduct the spatial
characterisation described in the previous subsection, i.e.\ we
identify and characterise all the states (i.e.\ dissipative
structures). We then, at each snapshot $m$, construct a `state map',
by which each structure in that snapshot is identified with a
structure in the previous snapshot, $m-1$. This is effected rather
crudely: a structure, $a$, in $m$ is identified with a structure, 
$b$, in $m-1$ if $a$ contains a grid cell that is in or adjacent to a
grid cell contained in $b$. Having done that, we can construct path
segments, with the path ending when the bijectivity of the state maps
breaks down. At these break-downs we can then identify which paths are
connected and construct `path maps', which permits us to identify
processes as the sum of connected paths. 

\subsubsection{Measurements}

The bijectivity of paths makes their
analysis substantially easier than the analysis of processes. 
The temporal evolution of the state parameters can be calculated   
$L(t), W(t),  T(t), V(t), A(t)$, and $D(t)$ over the duration of their
home structure
\begin{equation}
\tau=t^{\text{final}}-t^{\text{inital}}=(N_{s}-1)\Delta t
\end{equation} 
where $N_{s}$ is the number of constituent states,
 and $t^{\text{final}}$ and $t^{\text{initial}}$ are times at the final and initial
 states.
 From the energy dissipation rate the total energy dissipated by the path can be calculated
\begin{equation}
E=\int_{\text{path}}  D(t) dt=\sum^{N_{s}}_{k=1}D_{k}\Delta t.
\end{equation}
Another useful measurement is the peak of a state quantity along the path i.e.
\begin{equation}
X_{\text{max}}=\max_{\text{path}}\left[X(t)\right]=\max_{m\in \lbrace 1...N_{s}\rbrace}\left[\lbrace X_{m}\rbrace\right],
\end{equation}
where $X$ represents any of the measurements. These are however sensitive to chaotic fluctuations so must be used with care. 

Process properties are more difficult to calculate. 
Consider a given process with constituent paths with labels
$n=1,...N_{p}$, each with initial and final times $t^\text{initial}_{n}$
and $t^{\text{final}}_{n}$. 
Let $E_{n}$ be the total energy dissipated by path $n$ and $X_{\text{max},n}$ be the peak of a state quantity along path $n$. The duration of the process is given by 
\begin{equation}
\tau=\max_{n\in\lbrace 1...N_{p}\rbrace}\left[\lbrace
  t_{n}^{\text{final}}\rbrace\right]
-\min_{n\in\lbrace 1...N_{p}\rbrace}\left[\lbrace t_{n}^{\text{initial}}\rbrace\right].
\end{equation}
The total energy dissipated by a process is found by summing over the constituent paths
\begin{equation}
E=\sum^{N_{p}}_{n=1}E_{n}.
\end{equation}
 The peaks of the measurement parameters can also be calculated over the whole process 
\begin{equation}
X_{\text{max}}=\max_{n\in\lbrace 1...N_{p}\rbrace}\left[\lbrace X_{\text{max},n}\rbrace\right].
\end{equation}

\section{Results: spatial analysis}

\subsection{Intermittency}
 
\begin{figure}
\centering
\includegraphics[width=7cm]{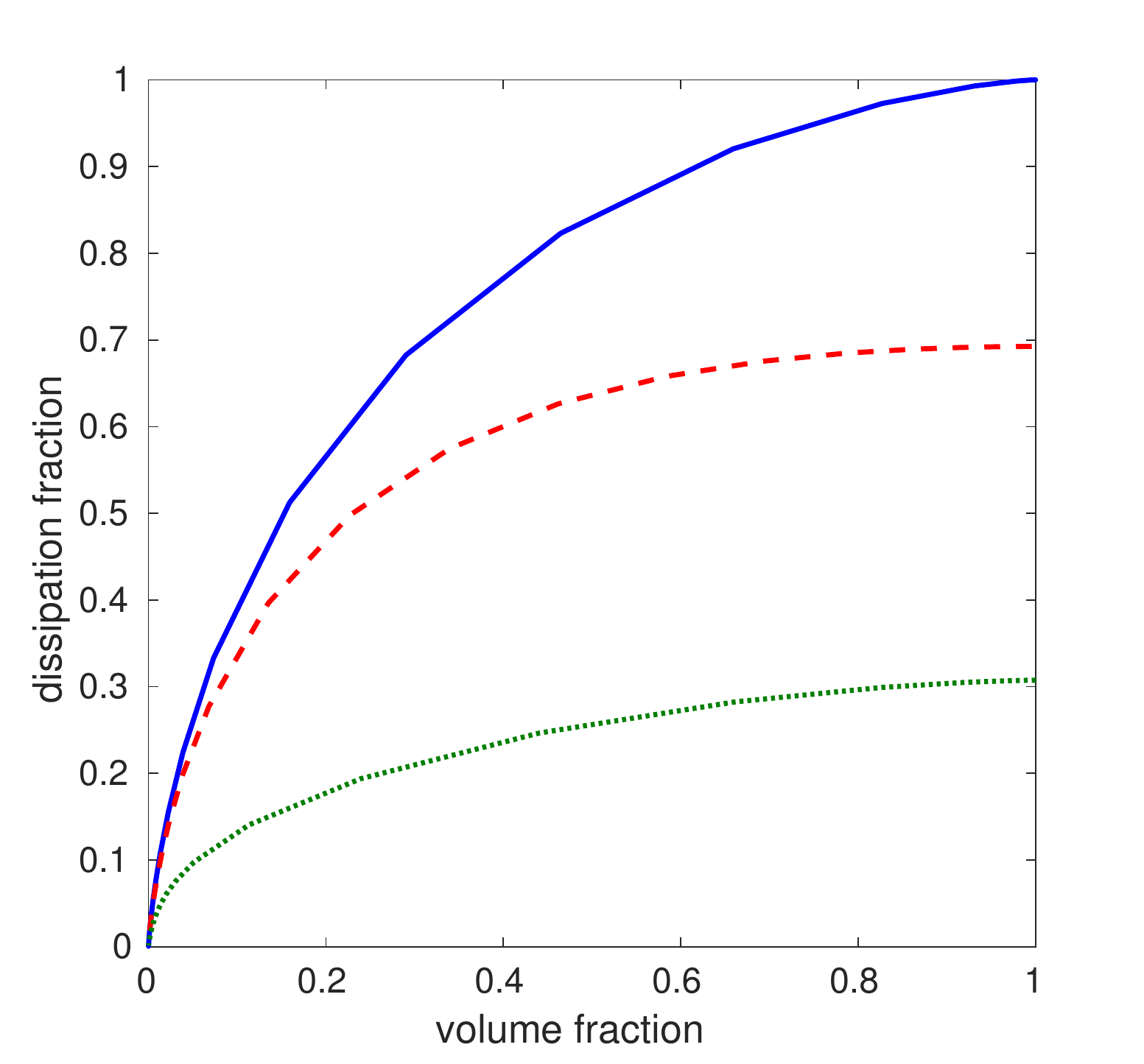}
\caption{Total dissipation fraction against volume fraction. Viscous
  dissipation is shown by the green dotted curve, the red dashed curve is
  Ohmic dissipation, and the sum of Ohmic and viscous is shown by the
  solid blue curve. The Reynolds numbers are $\text{Re}=1250$ and $\text{Rm}=5000$. 
The data is taken from simulation zn256a.}
\label{fig::znf256dfvf}
\end{figure}

Before we extract and characterise individual dissipative structures
we provide quantitative measures of their intermittent nature. As is well-known,
turbulent energy is thermalised primarily in narrow localised regions and 
in short bursts, not uniformly in space and time. In this
subsection we focus only on
spatial intermittency.

The dissipation fractions associated with Ohmic, viscous, and total
heating are plotted against volume fraction in Figure
\ref{fig::znf256dfvf}. The simulation in question
is znf256,  and the measurements were taken after at least 20 orbits, once
the turbulence had settled down into a steady state.
The fractions are calculated by fixing $k$,
determining the total dissipation and volume of all identified structures
(see Section 2.3), then iterating through a range of $k$ at a given
snapshot. Finally we average over multiple snapshots.

Figure \ref{fig::znf256dfvf} tells us that
approximately twice as much energy is dissipated by Ohmic
heating than by viscous heating, as is expected from MRI turbulence ---
partly because $B^{2}$ greatly exceeds
$u^{2}$. Moreover, the shapes of the Ohmic and viscous
dissipation profiles differ: Ohmic dissipation rises
rapidly at the origin before plateauing, while viscous dissipation
rises less steeply. Ohmic heating is thus more spatially intermittent: 
 $50\%$ of the Ohmic dissipation occurs within $10\%$ of
the volume. One
expects that as the Reynolds numbers increase, dissipation
becomes even more concentrated, and indeed at $\text{Rm}=\text{Re}=45,000$, ZWBL17
find that $50\%$ of Ohmic dissipation takes place in $7.5\%$ of the
volume. No doubt there exists a scaling relation connecting the latter
fraction with Rm (for given Pm), but this is
numerically inaccessible at the moment. Nonetheless, the main qualitative point to
take away is quite clear: dissipation in MRI turbulence is
inhomogeneous, and will potentially generate extremely narrow regions of
hotter gas floating in a much larger expanse of cooler gas. 

Finally we compared the total dissipation fraction curves produced by znf256
and the less well-resolved znf128 and found excellent agreement. This
indicates that grid dissipation contributes negligibly to the
formation and behaviour of identified dissipative structures (for the
Re and Rm chosen). However, in znf256
if we compare the total energy injection rate (via the
action of the turbulent stress on the Keplerian shear) against the
total Ohmic and viscous dissipation rate, we find a shortfall of
roughly $10\%$. We conclude that this anomalous heating arises 
primarily from grid dissipation in shocks. Because this dissipation route is
subdominant we do not address it further in this paper, but we note
that shock heating's contribution to the energy budget may
be important in the faster flows associate with strong
net magnetic fluxes.

\subsection{Current sheets}

We next look at individual structures. We primarily use
 the zero-net flux simulations znf256, and an extraction threshold $k=6$, similar to values 
 in previous studies (Zhdankin et al.~2015b, 2016)
In addition, we checked that our main results are relatively robust with respect to $k$, for
threshold values $3\lesssim k\lesssim8$, at least for the resolution and Reynolds numbers we use. 
Individual structures are extracted at $20$ distinct times (`frames') each separated by
an orbit. As most structures have lifetimes less
than an orbit this means we rarely count the same structure twice (and
even then if we have it will have evolved significantly from when it started).

\begin{figure}
\centering
\includegraphics[width=9cm]{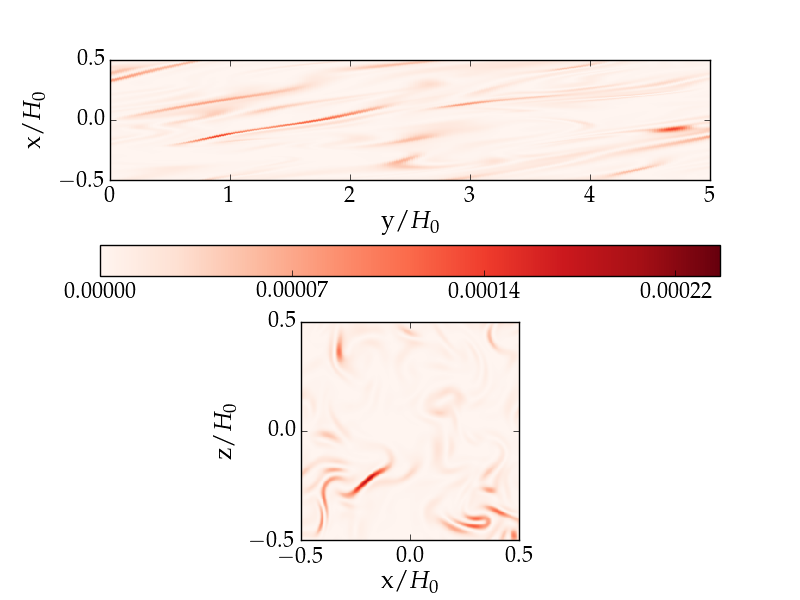}
\caption{Colour maps of $J^2$ in $xy$ and $xz$ planes from simulation
  znf256. These correspond to the Ohmic dissipation rate. As is
  clear, dissipation is concentrated into tilted and elongated ribbons
in the $xy$ plane. Their narrowness is also shown in the $xz$ plane
slice, but here have no readibly identifiable
orientation. }
\label{fig::identification}
\end{figure}

\begin{figure*}
\centering
\includegraphics[width=13cm]{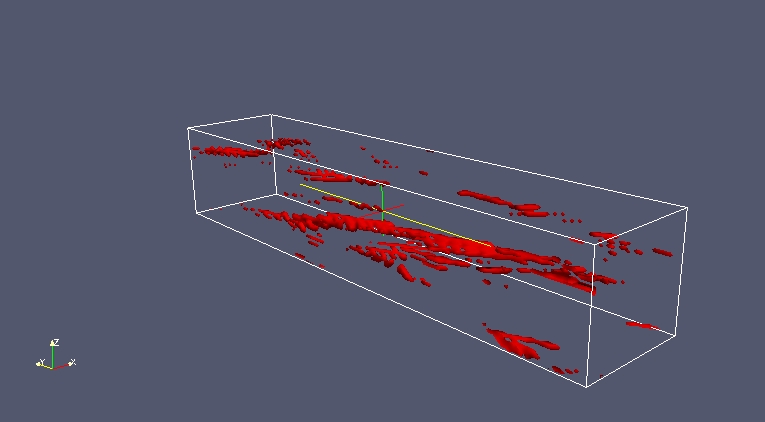}
\caption{ 3D rendering of Ohmic dissipative structures in znf256 with
  threshold set to $k=3$. Note that this $k$ is smaller than used for
  our qualitative analysis, but makes the structures easier to see by eye.}
\label{fig::strucutre}
\end{figure*}

To illustrate the results of the extraction, in Figure
\ref{fig::identification} we plot $J^2$, a proxy for $\epsilon_{\eta}$, along with
identified structures in $xy$ and $xz$ slices from znf256. 
A $3D$ rendering of these dissipating filaments and ribbons is shown in Figure \ref{fig::strucutre}.
(Note that in order to make them easier to see, these figures have been prepared with a smaller $k$.)
In the next few subsections we discuss several aspects of these features.

\subsubsection{Geometry}

A cursory glance at Figures \ref{fig::identification} and
\ref{fig::strucutre} show that the dissipative features are elongated and oriented 
along azimuth direction, but canted by some small angle. 
This is due, presumably, to shearing out by the differential rotation. 
Their geometry can be quantified through the vector
$\mathbf{\hat{L}}$.

\begin{figure}
\centering
\includegraphics[width=9cm]{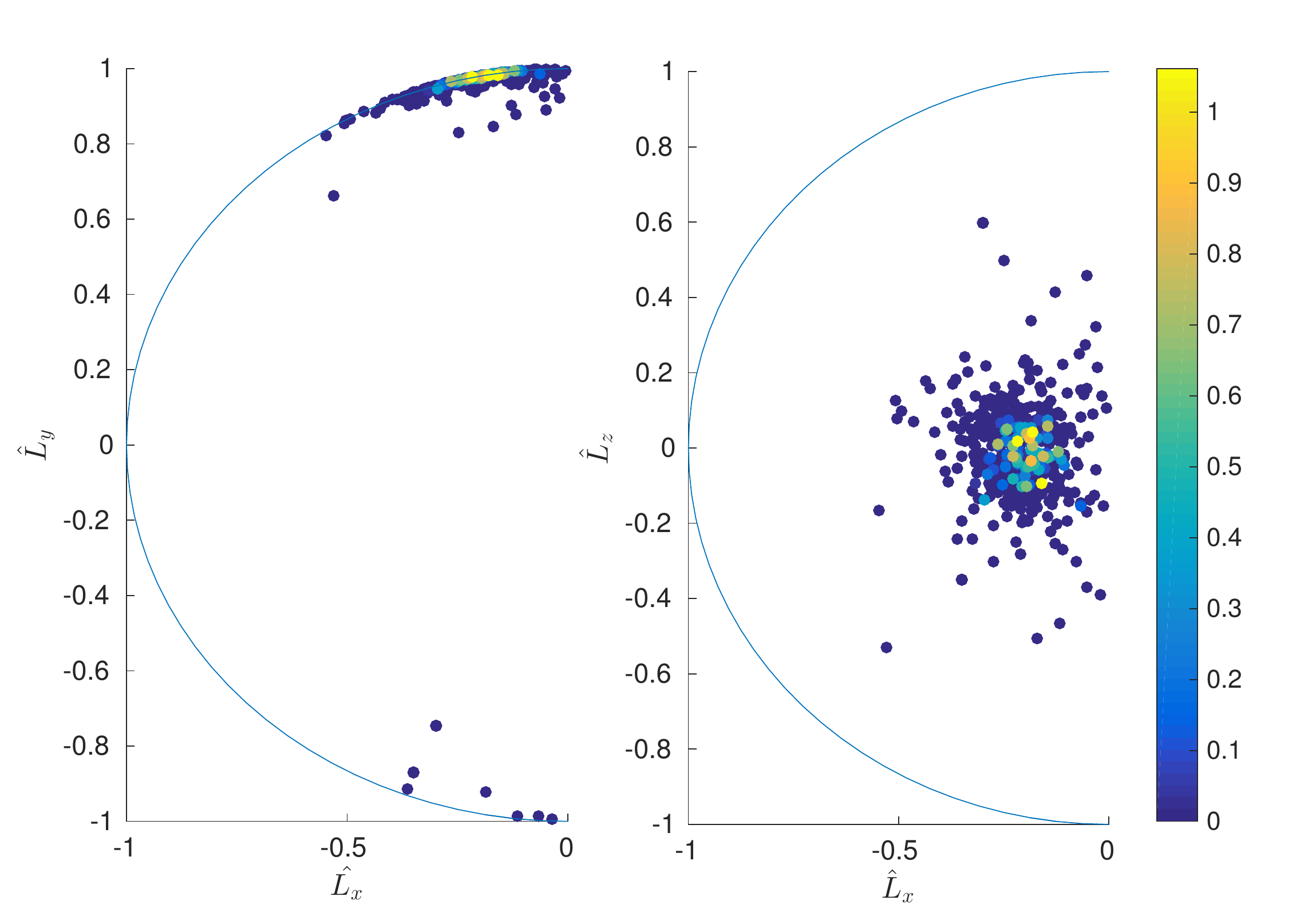}
\caption{ Projections of $\hat{\mathbf{L}}$ for multiple structures
  taken from znf256. The left panel reveals
 the characteristic tilt angle of the current sheets is
 $\theta\approx11.6^{\circ}$. 
The right panel shows no preferred orientation in the $xz$ plane.
The color of the dots indicate the length of the structure.}
\label{fig::L_projection}
\end{figure}

 In Figure \ref{fig::L_projection} we show the
projection of $\mathbf{\hat{L}}$ on the $xy$ and $xz$
planes for multiple structures. $\mathbf{\hat{L}}$ is found to be largely aligned along
$\mathbf{\hat{e}}_{y}$, with a small offset in
$\mathbf{\hat{e}}_{x}$. Their projections onto $\mathbf{\hat{e}}_{z}$
are typically small, especially  for the longest structures with
$L>0.1H$. 
We define a tilt angle, $\theta$, with respect to $\mathbf{\hat{e}}_{y}$
\begin{align}
\theta=\tan^{-1}\left(\frac{L_{x}}{L_{y}}\right)
\end{align}   
In our znf128 and znf256 simulations, we find
$\langle\theta\rangle\approx11.6^{\circ}$, where the average is taken over
all structures. 
This is consistent with, but slightly smaller than, the magnetic field
correlation tilt angle calculated by Guan et al.~(2009).  

\begin{figure}
\centering
\includegraphics[width=8cm]{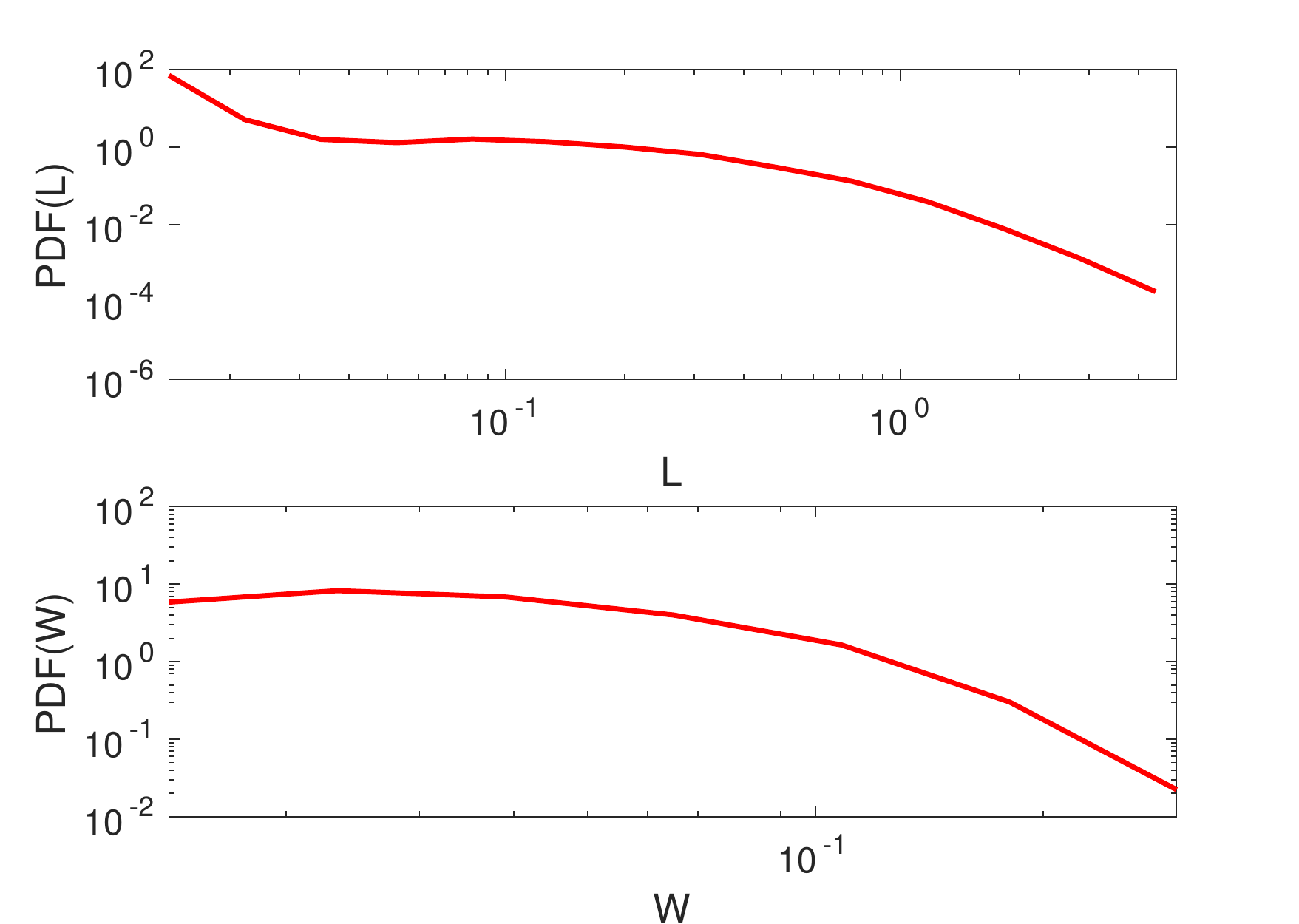}
\caption{Probability distribution functions of $L/H$ and $W/H$ for
  identified structures in simulation znf256.}
\label{fig::LWpdf}
\end{figure}

In Figure \ref{fig::LWpdf} we plot probability distribution functions
(PDFs) 
of $L$ and $W$. These curves
are dominated by a great many small structures that emerge and
disappear rapidly. However, there is a distinct plateau in both curves
at lengths above the grid scale. Mean values for $L$ and $W$ are
$0.696H$ and $0.097H$ respectively. The former is well separated from the
azimuthal box size $5H$ (the relevant numerical outer scale) and the physical
diffusion lengths, whereas the latter is
well separated from the grid $0.0039H$, and probably controlled by the
diffusivities. The large average length of the
dissipative structures is remarkable. It indicates that dissipation in
MRI turbulence occurs in regions that are meso-scale, of order
or less than the disk scale height. These ribbon-like features possess
widths and thicknesses controlled by the diffusivities, and in reality
are extremely thin, but at the same time are
 significantly elongated in azimuth. 
The fact that the mean $L$ is well
separated from the outer and diffusion scales gives us hope that this
basic result is independent of the numerical particulars of our simulation.

 Figure \ref{fig::AL_znf256a}b plots $W$ against $L$ for all
 extracted structures in znf256. 
While $W\ll L$, there appears to be an approximate linear correlation
between the two variables: 
the longer the structure, the wider the structure. 
Next, If we assume that the dissipative structures are in the form of sheets, with length $L$, width $W$ 
and thickness $T$ and take $W,L\gg T$, then the area of a structure
will be approximately given by $A\approx 2WL \propto L^2$. 
In Figure \ref{fig::AL_znf256a}c we plot $A$ against $L$ and indeed find
there to be a linear correlation between $A$ and $L^{2.2}$. For smaller $L$,
the ends of the structures give a non-negligible contribution to
$A$.

In contrast to $L$ and $W$, there is no clear scaling between the
thickness $T$ and the length or width, as Figure \ref{fig::AL_znf256a}d
makes clear.
 The range in thickness is also  smaller than the other dimensions,
 though still roughly an order of magnitude greater than the grid.

\subsubsection{Dissipation}

\begin{figure*}
\centering
\includegraphics[width=15cm]{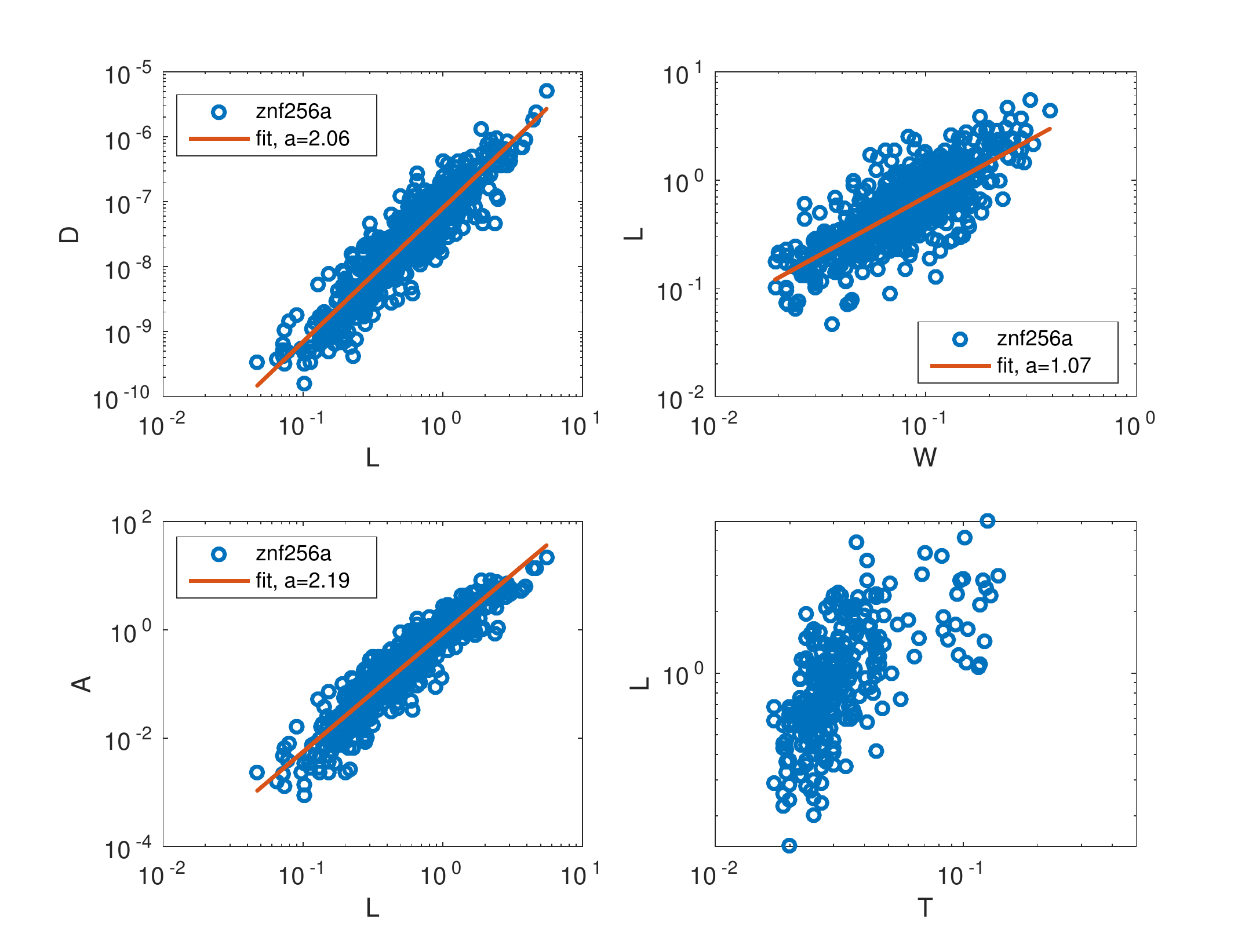}
\caption{Correlations between various structure measurements along
  with best fit lines (power laws with exponent $a$), when appropriate, from simulation
  znf256. Structures that have $W<4\delta x$ are discarded for all
  scatter plots and for the $L$ vs $T$ scatter plot we include the
  constraint $T>4\delta x$. }
\label{fig::AL_znf256a}
\end{figure*}

In Figure \ref{fig::AL_znf256a}a we plot the power dissipated by a
structure against the length of the structures, and this reveals a
strong correlation between the two measurements with  approximately $D\propto L^{2}$. 
If we assume the power dissipation is \emph{uniformly} distributed within a
structure, 
then this scaling can be achieved
using a similar argument to earlier. 
The larger power dissipated in 
larger structures is purely a geometrical effect: 
larger structures dissipate more not because they are more intense, 
but because they are simply larger. This is another important point uncovered by our simulations.

\begin{figure}
\centering
\includegraphics[width=8cm]{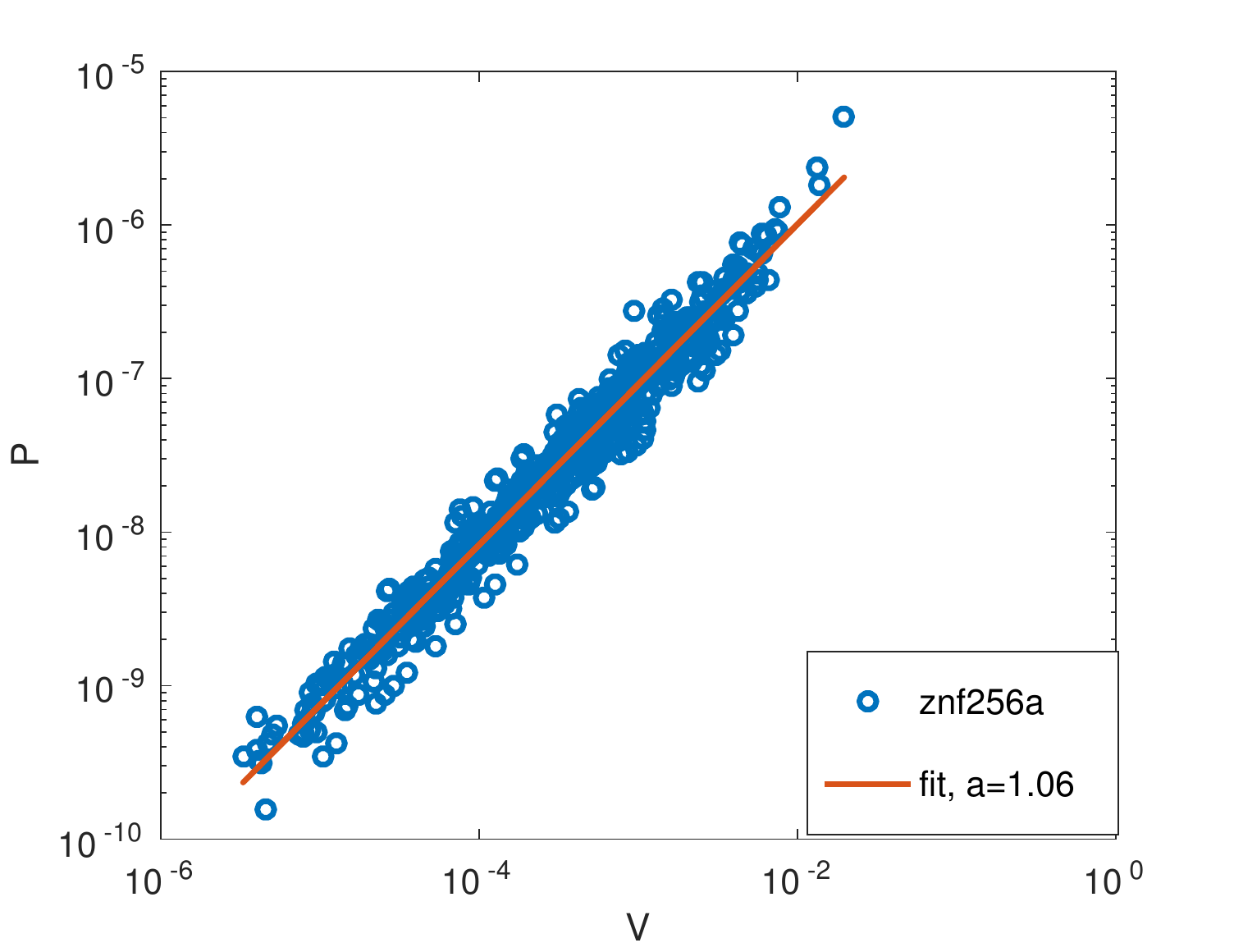}
\caption{Magnetic dissipation in a structure as a function of its
  volume in simulation znf256. Superimposed is a line of best fit
  $D\propto V^a$, with $a=1.06$. }
\label{fig::PV_Re}
\end{figure}

To further illustrate this, in Figure \ref{fig::PV_Re} we plot the magnetic dissipation rate against 
the volume of the structures. The power dissipated is found to be
proportional to volume,
 i.e. the dissipation rate per unit volume is nearly a constant, as expected. This constant is 
dependent on the energy of the flow and the Reynolds numbers. We
calculate  
$\langle D/V\rangle$ to be $1.44\times10^{-4}$ 
for znf256, where the angled brackets denote averaging over all current sheets.

The scalings obtained from our highest resolution simulations appear
to be compatible with those associated with incompressible, nonshearing, nonrotating, forced RMHD
simulations (Zhdankin et al.~2016) even though our simulations possess no imposed
external magnetic field and the flow is compressible. In the MRI
simulations, however, the shearing motion stretches any radial field into a
strong toroidal field which can then play a similar role to the
guiding field in RMHD. It is also worth noting that isothermal zero
net-flux MRI simulations are only mildly
compressible with only $\lesssim10\%$ variations in density. These
results suggest that dissipation in MRI turbulence is not special to
the MRI —- the features are comparable to other MHD systems. 
A similar conclusion was reached by Walker et al.~(2016).

\begin{figure}
\centering
\includegraphics[width=9cm]{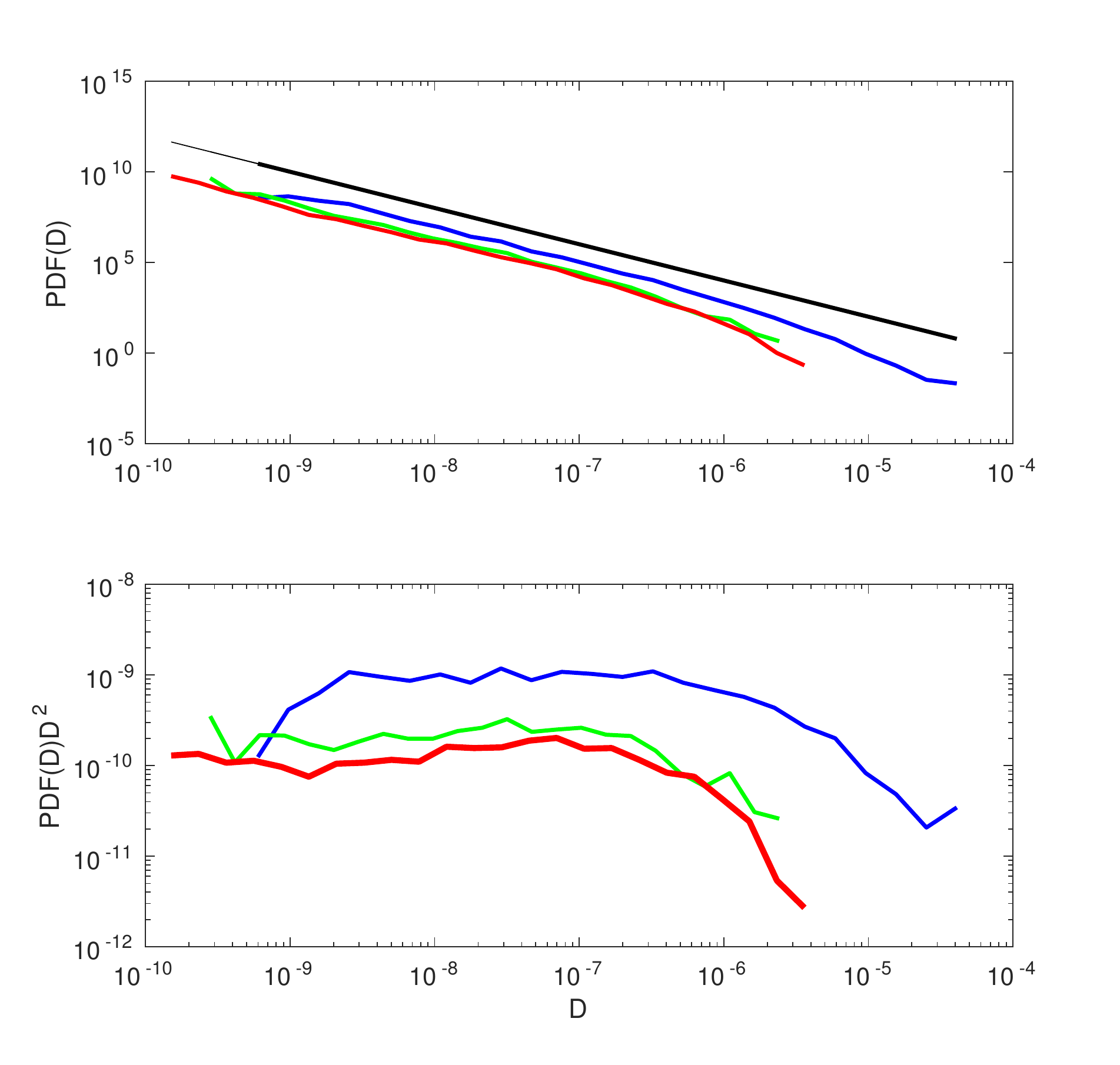}
\caption{Top panel: Probability distribution of the magnetic energy
  distribution rate $D$ for znf256 (red), vert128 (blue), and
  tor128 (green). For reference we impose a power
  law curve with the critical index $-2$ (black).
  Bottom panel: as in the top panel but multiplied by $D^{2}$.}
\label{fig::PDFp}
\end{figure}

Zhdankin et al.~(2014) observed that the probability distribution of
the dissipation rate of structures in MHD turbulence tends
towards a power law with an index of $-2$ as the Reynolds numbers are
increased. Remarkably, this value is the critical index at which weak
and strong structures contribute equally to the total energy
dissipation (Hudson 1991): the relatively large number of weakly
dissipating structures balances out the fact that they are only weakly
dissipating, hence they make a contribution to the total energy
budget that is comparable to the strongly dissipating
structures. In Figure \ref{fig::PDFp}a we plot the probability
distribution of $D$ for znf256 along with a
curve with the critical index. For an intermediate band of $D$, the
probability distribution function is compatible
with a power law distribution with an index close to $-2$. In Figure
\ref{fig::PDFp}b, we plot the probability distribution function
multiplied by $D^{2}$, to better reveal the scaling. Note that there is a drop off in power at
high $D$. It is possible that this is due to insufficient statistics
for these infrequent strongly dissipating events. It is evident,
nonetheless, 
 that current sheets with a considerable range in 
dissipation rates contribute significantly to the total magnetic energy dissipated.

\begin{figure}
\centering
\includegraphics[width=8cm]{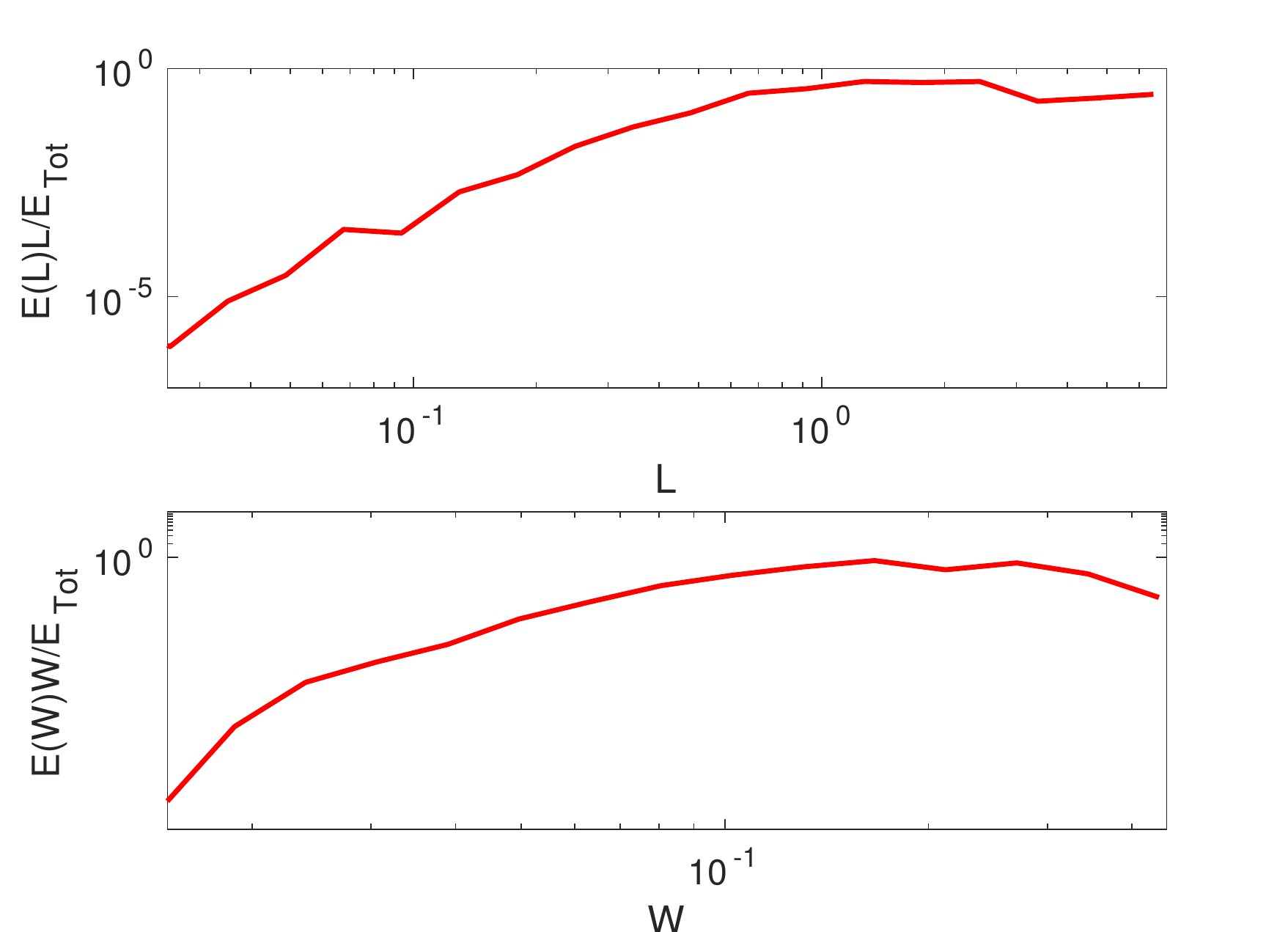}
\caption{Compensated energy dissipation rate $E(X)X$ for $X=L$ and $W$
  normalised by the total magnetic energy dissipated, for  znf256 }
\label{fig::compensated}
\end{figure}

Finally, the compensated energy dissipation rate provides a means of
determining upon which spatial scale the energy is being dissipated. 
Let $E(X)dX$ be the total energy dissipation rate for structures
with scales between $X$ and $X+dX$ where $X$ is any characteristic
scale. Then the maximum of the compensated energy dissipation rate,
$E(X)X$, gives the value of $X$ at which most of the energy
dissipation occurs. In Figure \ref{fig::compensated} we plot this
function for $X=\lbrace L, W\rbrace$. First, 
 consider $E(L)L$. There is a tendency for the energy to be dissipated at the longest of scales, $L\sim H$.
A similar trend is also suggested by $E(W)W$. 
Zhdankin et al.~(2016) also observe this feature
in their RMHD simulations. If our results generalise to higher
Reynolds numbers, dissipation is not necessarily on small scales as
one might expect. If strongly dissipating structures possess $L\lesssim
H$, they may interact with other
large-scale elements of the flow.

\subsection{Net-flux simulations}

It is worth asking whether there is a notable difference 
 between simulations with a net field and those
without. Here we attempt to explore the question by considering setups with
imposed 
vertical and toroidal fields. To
suppress strong channel flows we use a relatively modest $\beta$ for
the net-vertical simulation $\beta=400$, whereas in the toroidal field  
 simulation we set $\beta=200$. The resolution in both is $n=128$.

We find the inclination angle of the toroidal field simulation to be
$\theta=11.1^{\circ}$ which is comparable to the zero net flux
simulations, however the vertical field simulation has a lower value,
$\theta=9.4^{\circ}$. The latter contrasts with ZWBL17 who
 obtained a larger tilt
angle of $<\theta>=17.5^{\circ}$. 
Either the selection criteria for choosing their
structures differs to ours
 or, possibly, $\theta=\theta(\text{Re},\text{Rm},\beta)$.

We calculate the dissipation rate probability distribution function
for the two net-flux simulations and an identical zero-net-flux
simulation. In all three simulations the power law index of the
distribution is approximately $-2$ for over two decades in $D$, as
shown in Fig.~8. Hence
the energy is dissipated evenly over this range, which is in good
agreement with ZWBL17.

For this modest choice of plasma $\beta$, the presence of an imposed
field appears to only have minimal impact on the current
sheets. Instead their properties are largely determined by the
dissipative coefficients and the shear. Zero-net-flux simulations
generate their own local net $B_{y}$ which works like an
imposed field, as far as dissipative structures are concerned and at least
for structures not too elongated. If a strong net vertical field is
imposed, then differing results are
expected due to the strong channel modes that develop.

\section{Results: temporal analysis}

We now address a number of questions regarding the structures'
evolution over time. How long, on average, do they persist and how
does this depend on their geometric properties? How do these
properties evolve over time and within a given path? How does the
dissipation rate vary over time? And finally, what degree of dynamical
interaction takes place between nearby structures?

 This task is considerably more computationally challenging than the
 spatial analysis,
 primarily due to the large amount of data to be processed when a high
 frame cadence is used. As a consequence, very few studies exist, the
 most notable being Zhdankin et al.~(2015b) for forced incompressible
 MHD, and Yang et al.~(2017) for forced compressible MHD.
 In order to make any progress,
 we present a limited survey using lower resolutions than earlier,
 but yet sufficient to make some ground-level points.
 Our resolution in this section is $n=64$, mainly using simulation
 znf64, and the time step between outputs is $\Delta t = 0.0032t_\text{orb}$
 taken over an interval of size $7.5t_\text{orb}$, where $t_\text{orb}$ is an
 orbital period. This interval is taken some 20 orbits after the simulation
 begins, so as to ensure that we have reached a statistically steady state.

\subsection{Individual processes}

\begin{figure*}
\centering
\includegraphics[width=7cm]{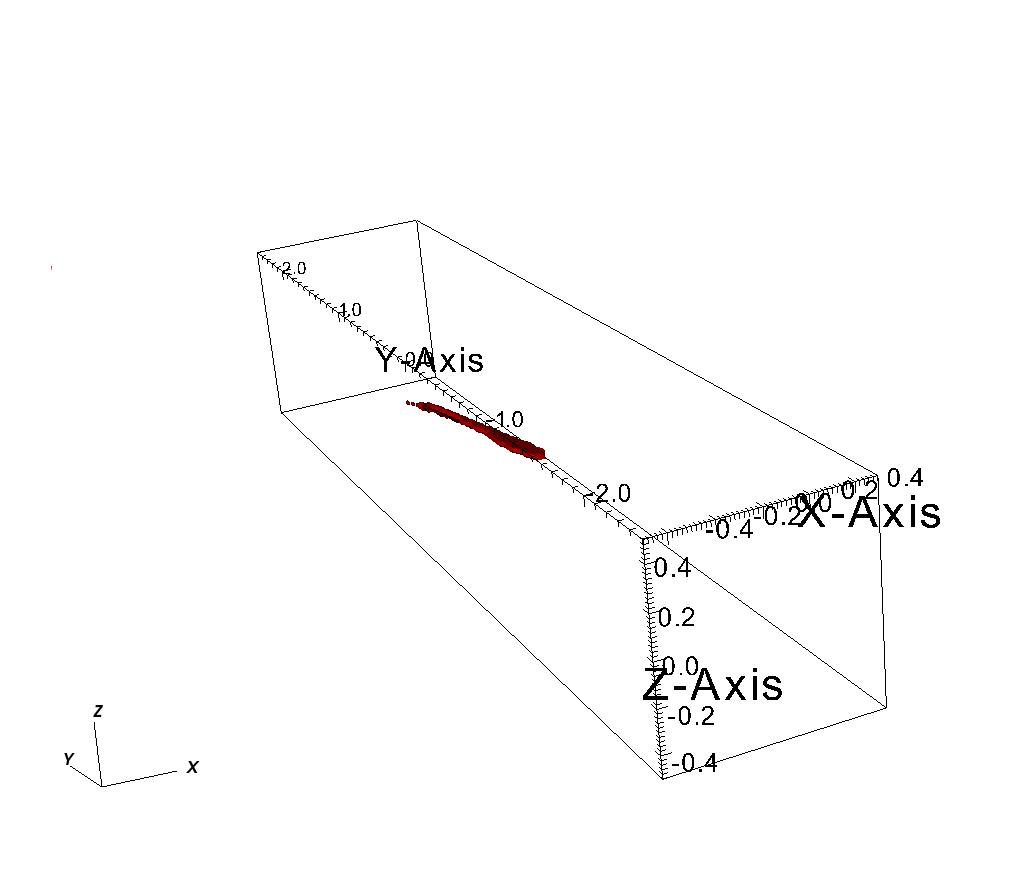}
\includegraphics[width=7cm]{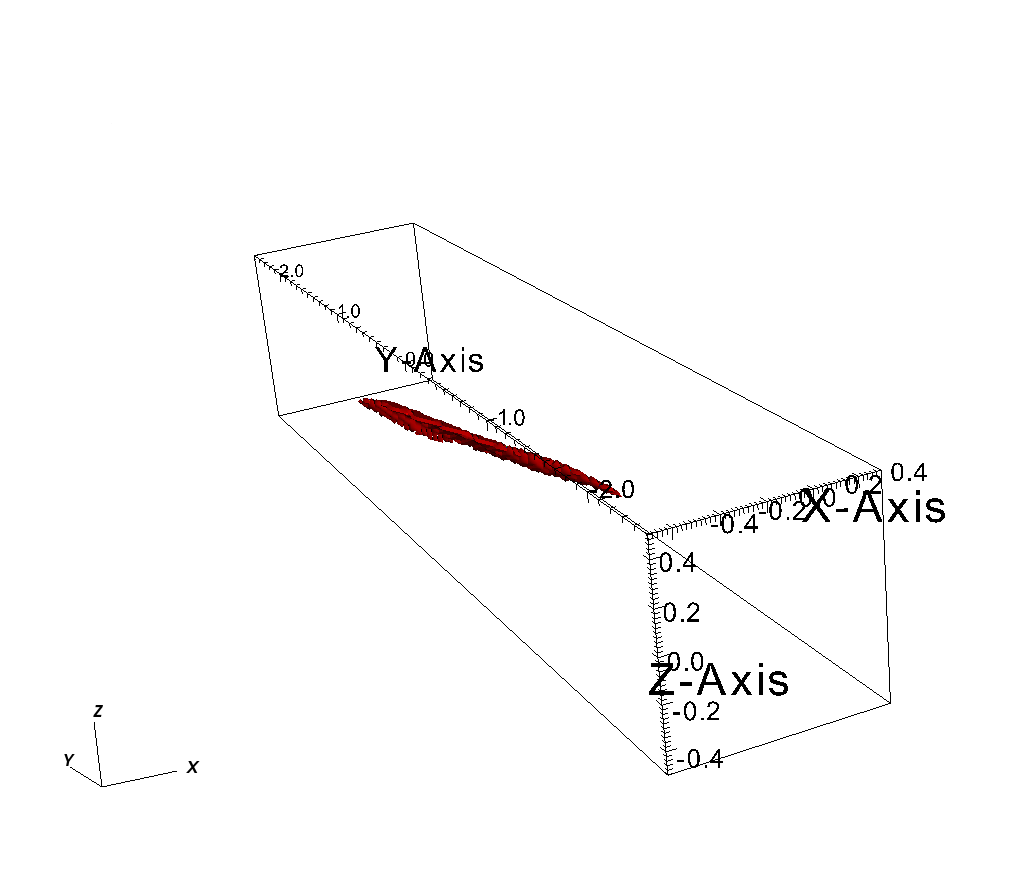}
\includegraphics[width=7cm]{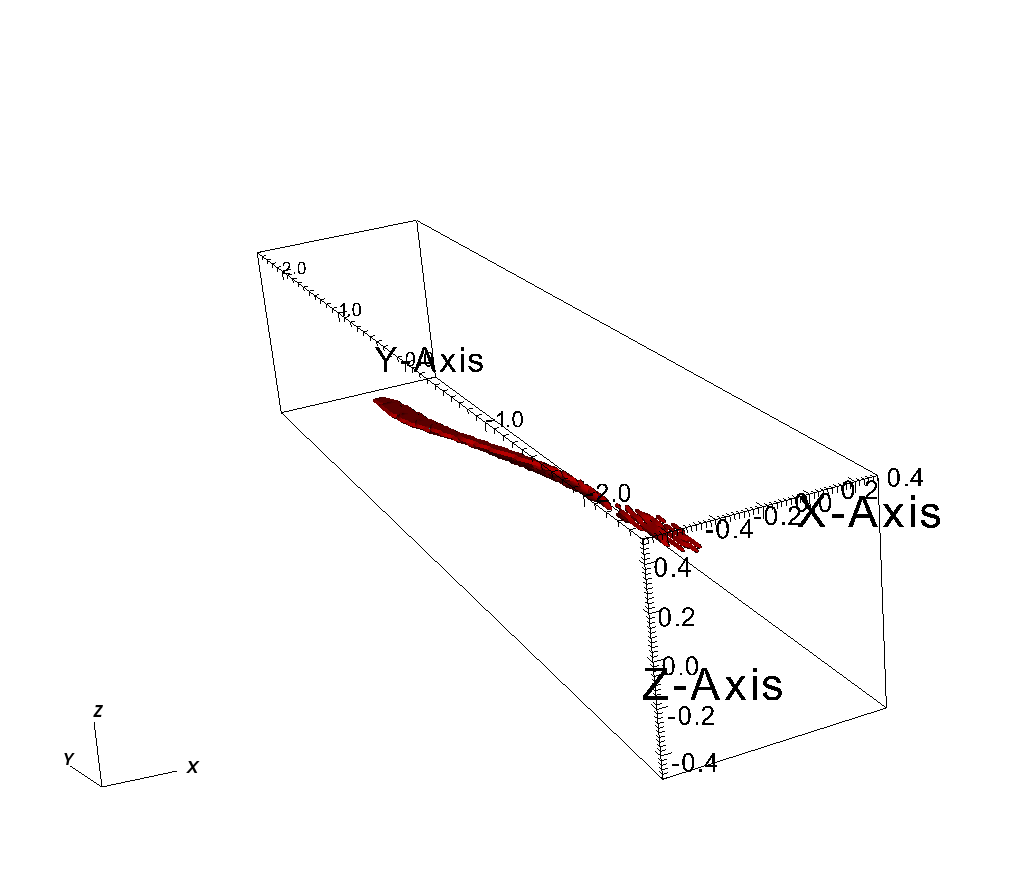}
\includegraphics[width=7cm]{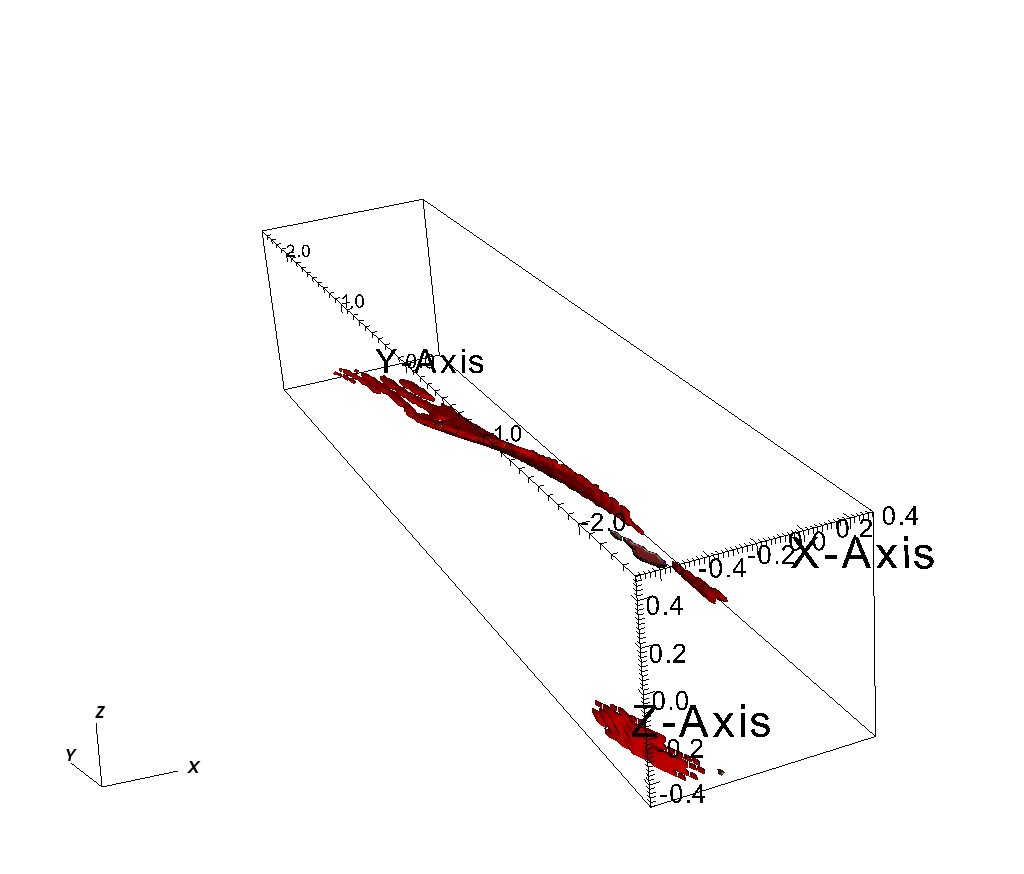}
\includegraphics[width=7cm]{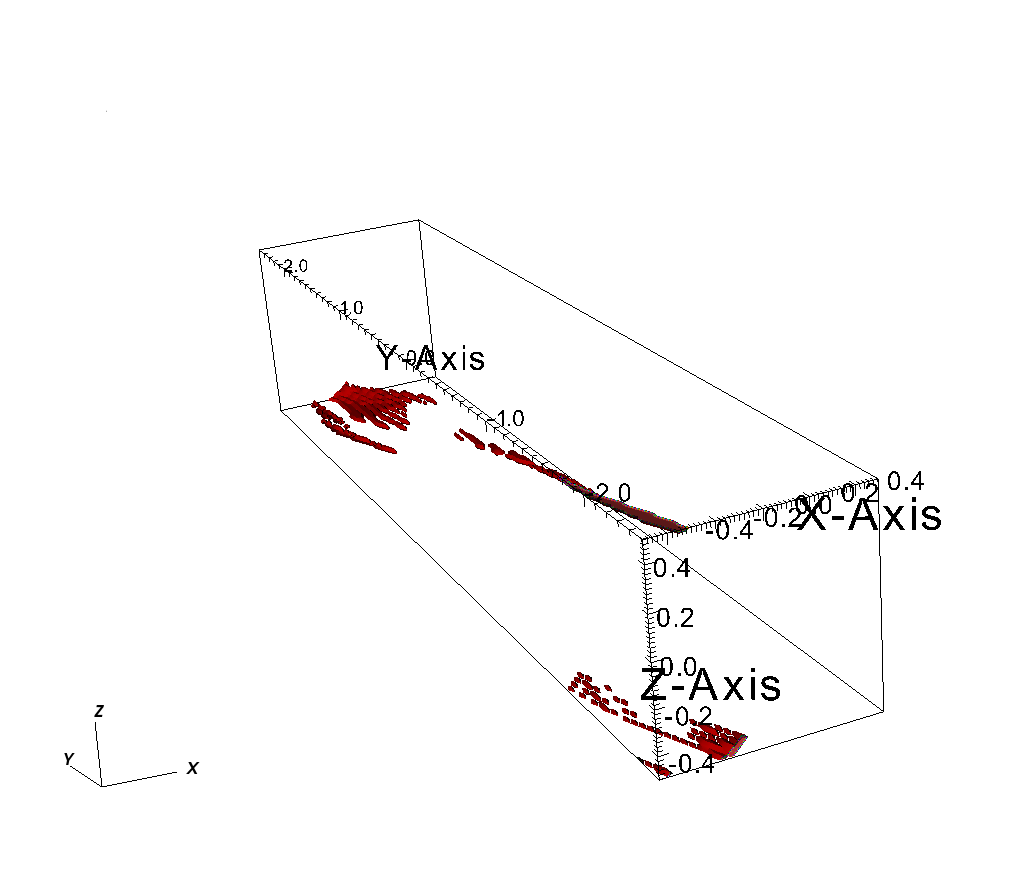}
\includegraphics[width=7cm]{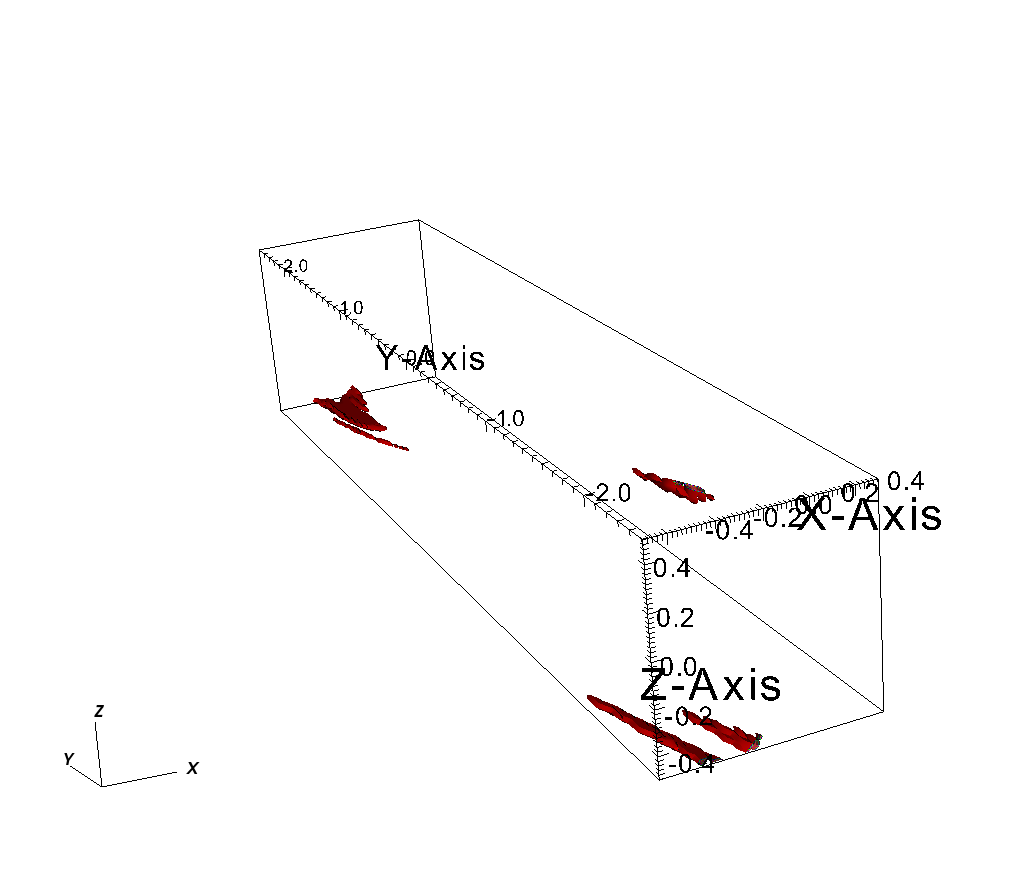}
\caption{Volume renderings of the biggest and longest-lived process in
  znf64 at
  $t=0.2$, $0.35$, $0.67$, $0.83$, $0.85$ and
  $0.98t_\text{orb}$. }
\label{fig::volrendtemp}
\end{figure*}

\begin{figure*}
\centering
\includegraphics[width=16cm]{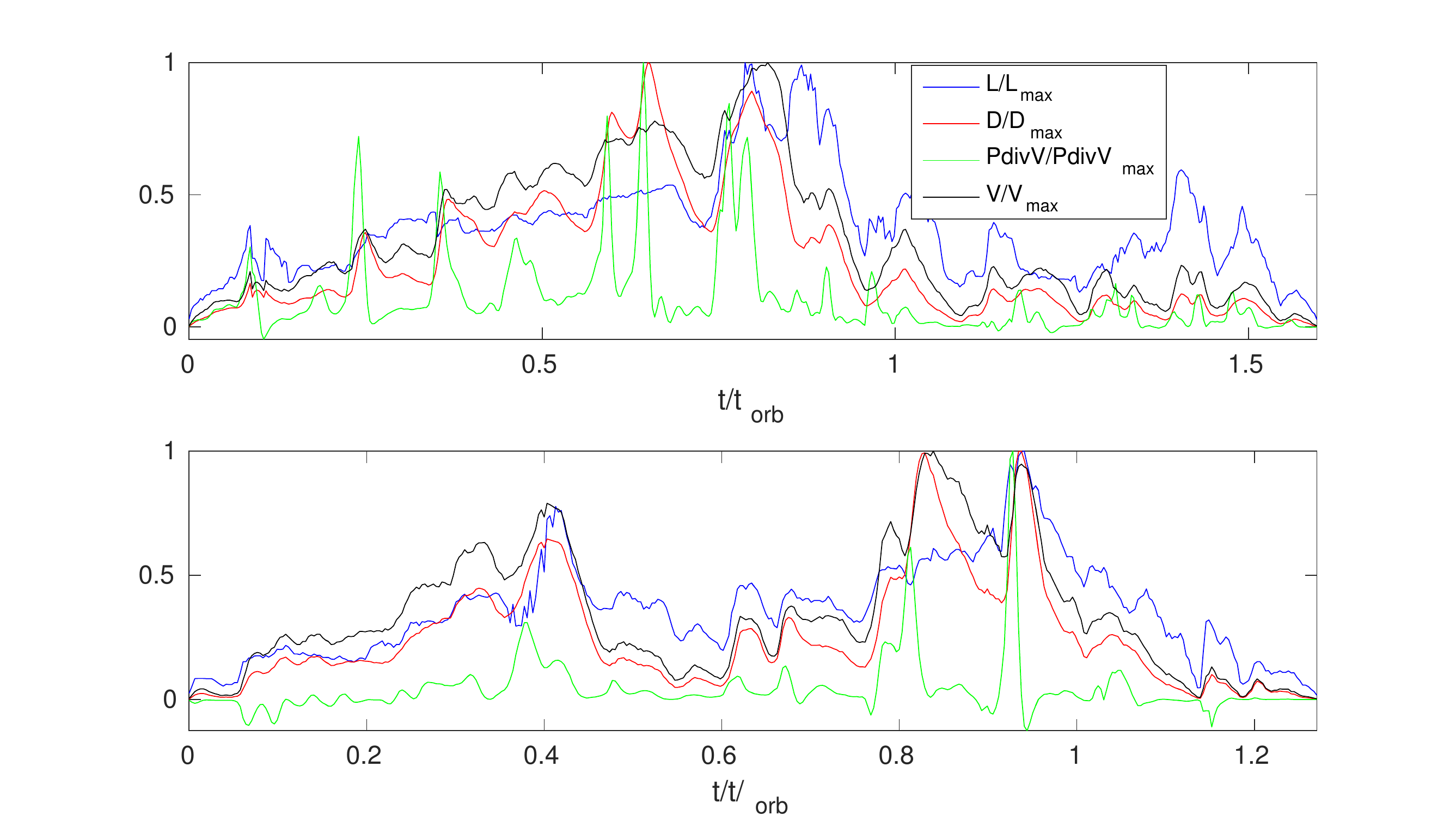}
\caption{Time evolution of various properties associates with two of
  the longest-lived processes in simulation znf64. Here $L$ is length,
$D$ is rate of dissipation, $P\nabla\cdot\mathbf{v}$ is pressure work,
and $V$ is volume. Each is normalised by its peak value.}
\label{fig::znf64Time}
\end{figure*}

First we concentrate on the longest processes, which typically survive
for roughly an orbit and a half.
Figure \ref{fig::volrendtemp} presents 3D snapshots describing the
evolution of one such process, composed of its various paths merging,
bifurcating, and disappearing. In Fig.~\ref{fig::znf64Time} we plot
the time evolution of $L$, $D$ and $V$ for the
longest two processes. As one can see, over the course of their
lifetimes these features grow in size, and in dissipative intensity,
before shrinking, disassociating, and ultimately
evaporating. Superimposed upon this general trend are short scale
variations, some of which we attribute to acoustic waves (see Section
5). 
The three
properties, $L,D$ and $V$, track each other fairly well, although there
are slight deviations. In particular, the volume and the 
power dissipation rate of the
processes evolve together, which indicates that the
dissipation rate per unit volume is nearly constant throughout the
duration of the process. 
This is consistent with the linear $D-V$ relation that 
we found in the spatial statistics (cf.\ Figure \ref{fig::PV_Re}).

Given the longevity of the most prominent processes, substantial energy can be
dissipated during their lifespans, and in optically thick media, where radiative transfer is
inefficient, this could induce considerable temperature inhomogeneity
in the plasma. In principle, the heating rate per unit volume can be
compared to an estimate of the cooling rate via radiative diffusion
across the thickness of the current sheet so as to estimate the
temperature change within the sheet. Of course, such estimates will be strongly
influenced by the thickness of the current sheets, which is
determined by $\eta$; as a consequence, they will be difficult to apply to realistic
systems that are far less resistive than the simulations.
 

\begin{figure*}
\centering
\includegraphics[width=12cm]{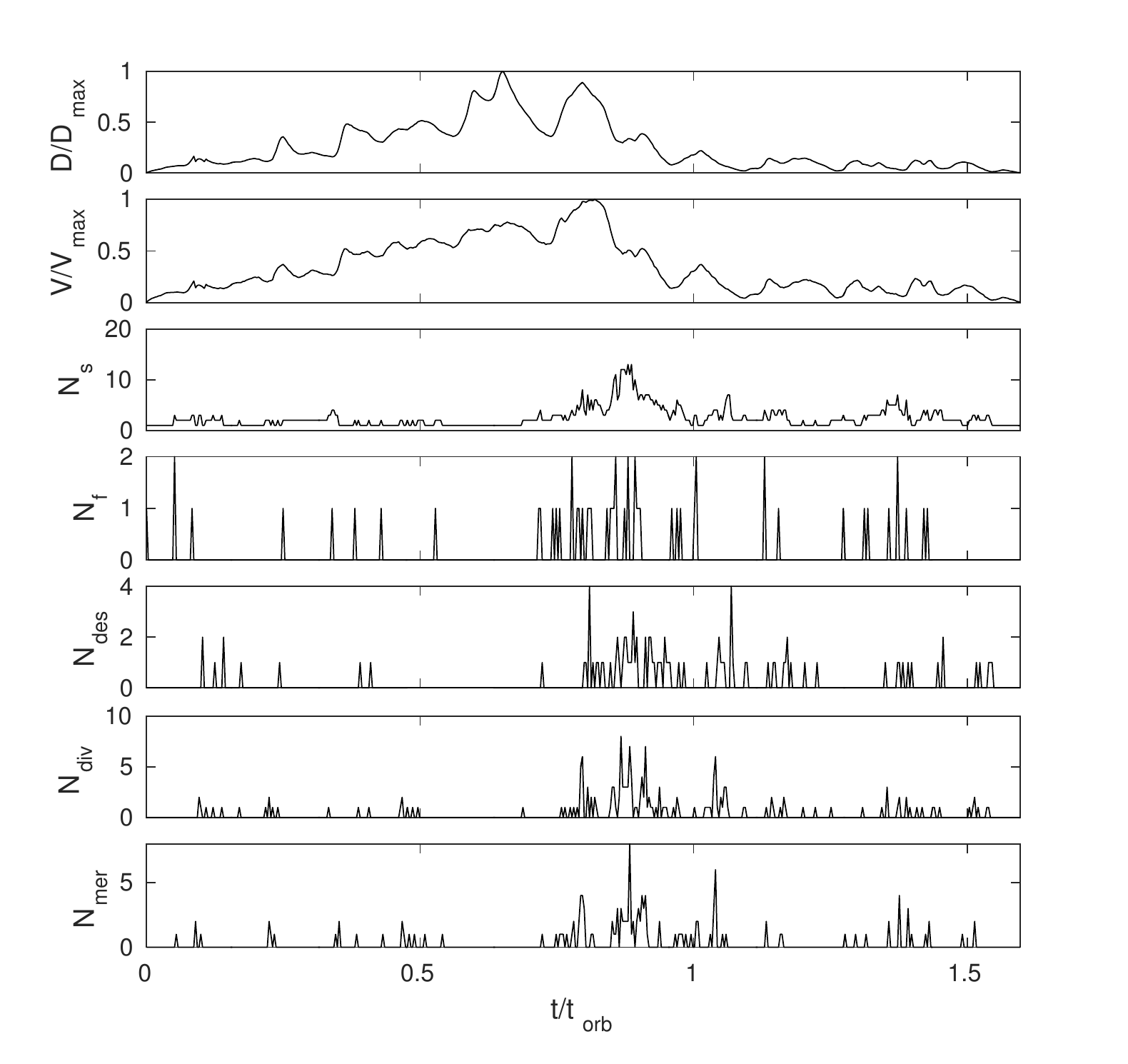}
\caption{Coherence history of the largest and longest-lived process
  from znf64. $N_{s}$ is the instantaneous number of structures,
  $N_{f}$ the number formations, $N_{des}$ destructions, $N_{div}$ divisions, and $N_{mer}$ mergers. }
\label{fig::znf64N}
\end{figure*}

We next explore the dynamical evolution of the component paths in a
given process. How many are they, and how do they interact?
Figure
\ref{fig::znf64N} shows the number of: structures $N_{s}$,
formations $N_{f}$, destructions $N_{des}$, divisions $N_{div}$,
and mergers $N_{mer}$, as functions of time for the longest process.
The volume of and total dissipation in the process is also included. 
We note that the change in $N_s$ must be equal to $\Delta
N_s=N_f-N_\text{des}+ N_\text{div} - N_\text{mer}$. 

 For the first $\sim0.75t_\text{orb}$ the process is relatively simple:
it is composed of only a handful of structures ($1-4$), though these
are growing in size and thus bringing about a growth in the total volume. During this
phase there is a general increase in the dissipation rate. At around
$t\approx 0.8 t_\text{orb}$, near the peak heating rate,
there is a sudden increase in complexity, with
 $N_{s}$ spiking at $12-23$ structures. The
fifth panel in Figure \ref{fig::volrendtemp} illustrates this
phase. There is a great deal of rapid formation, destruction,
division, and merging at this point, indicative of a climactic 
alteration in the morphology of the process. From this moment the
process steadily dies away.  This `temporal asymmetry' is also
witnessed in forced MHD (Zhdankin et al.~2015b), and  
it is tempting to attribute the
catastrophe at $t=0.8 t_\text{orb}$ to an instability of some form,
possibly of tearing or Kelvin-Helmholtz type (Loureiro et al.~2007,
2013, Samtaney et al.~2009). Regarding the classical tearing/plasmoid
instabilities, it would appear
that the observed Lundquist numbers and aspect ratios in our
simulations are too low for
their excitation,
but it is possible in our shearing, rotating, turbulent environment
these thresholds might be relaxed (see Section 6.2 in
Zhdankin et al~2015b). Certainly further work is required to nail down what is
going on here, but it is clear the tools of the temporal analysis (appearing in Fig.
\ref{fig::volrendtemp})
could provide an insightful way to study instabilities
and reconnection in turbulent current sheets.

\subsection{Statistics and scaling relations}

\begin{figure}
\centering
\includegraphics[width=7cm]{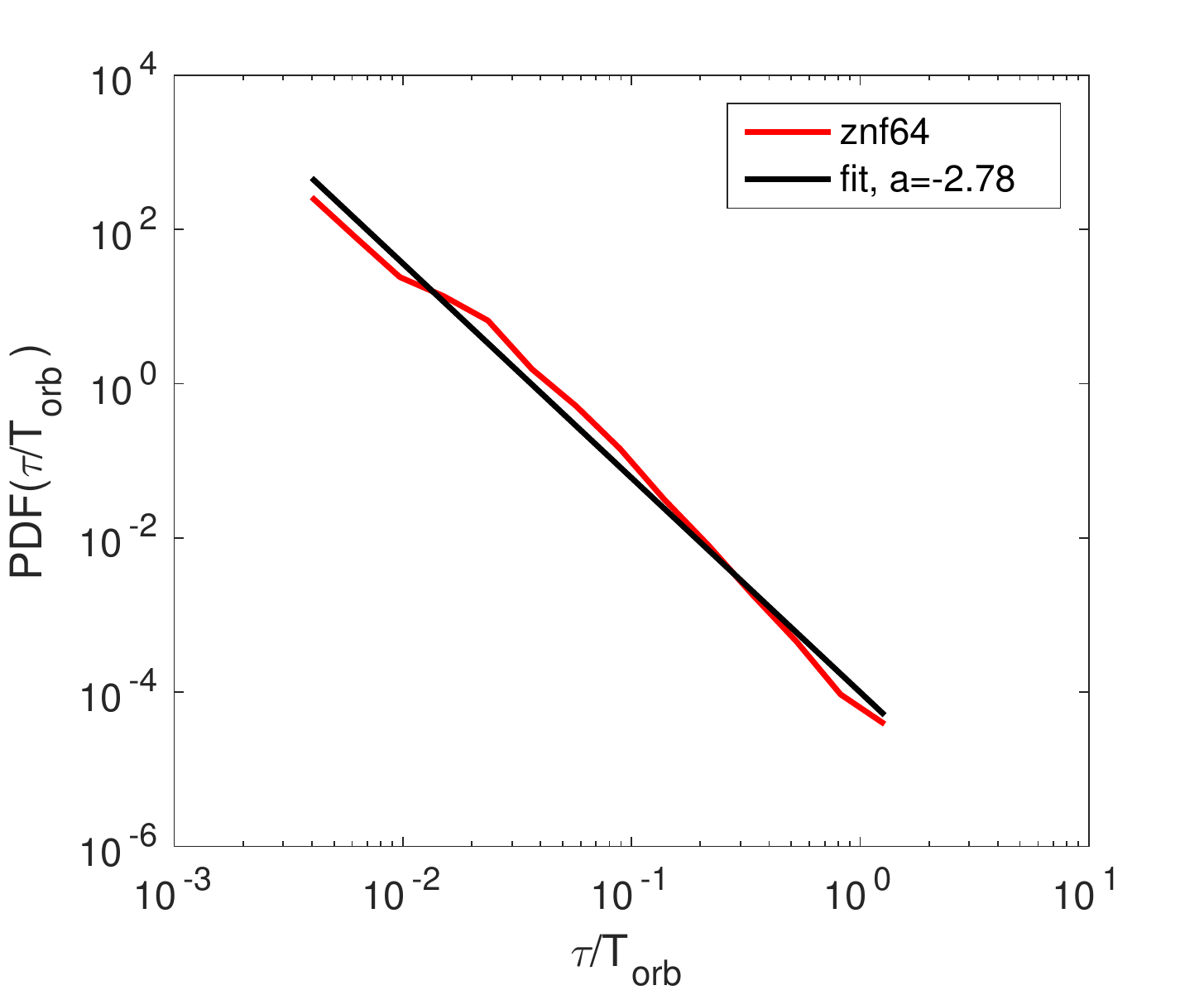}
\caption{Probability distribution function of process time $\tau$ in
  red. In black we have fitted a power law of exponent $-2.78$. }
\label{fig::znf64PvTa}
\end{figure}

In the previous subsection we focussed on only two long-lived
processes, both enduring for longer than one orbit. But what is the 
distribution of process lifetimes $\tau$?  In the lower
panel of Figure \ref{fig::znf64PvTa} we plot the lifetime probability
distribution function, only including complete
processes. The PDF is fitted by a power
law distribution with $\tau^{-2.78}$. Clearly long lasting processes
are in the minority, with the mean only some 1/100 of an orbit. 
That being said, during the $7.5t_\text{orb}$
analysed, we still detect $10$ processes that last longer than half an
orbit.

\begin{figure}
\centering
\includegraphics[width=9cm]{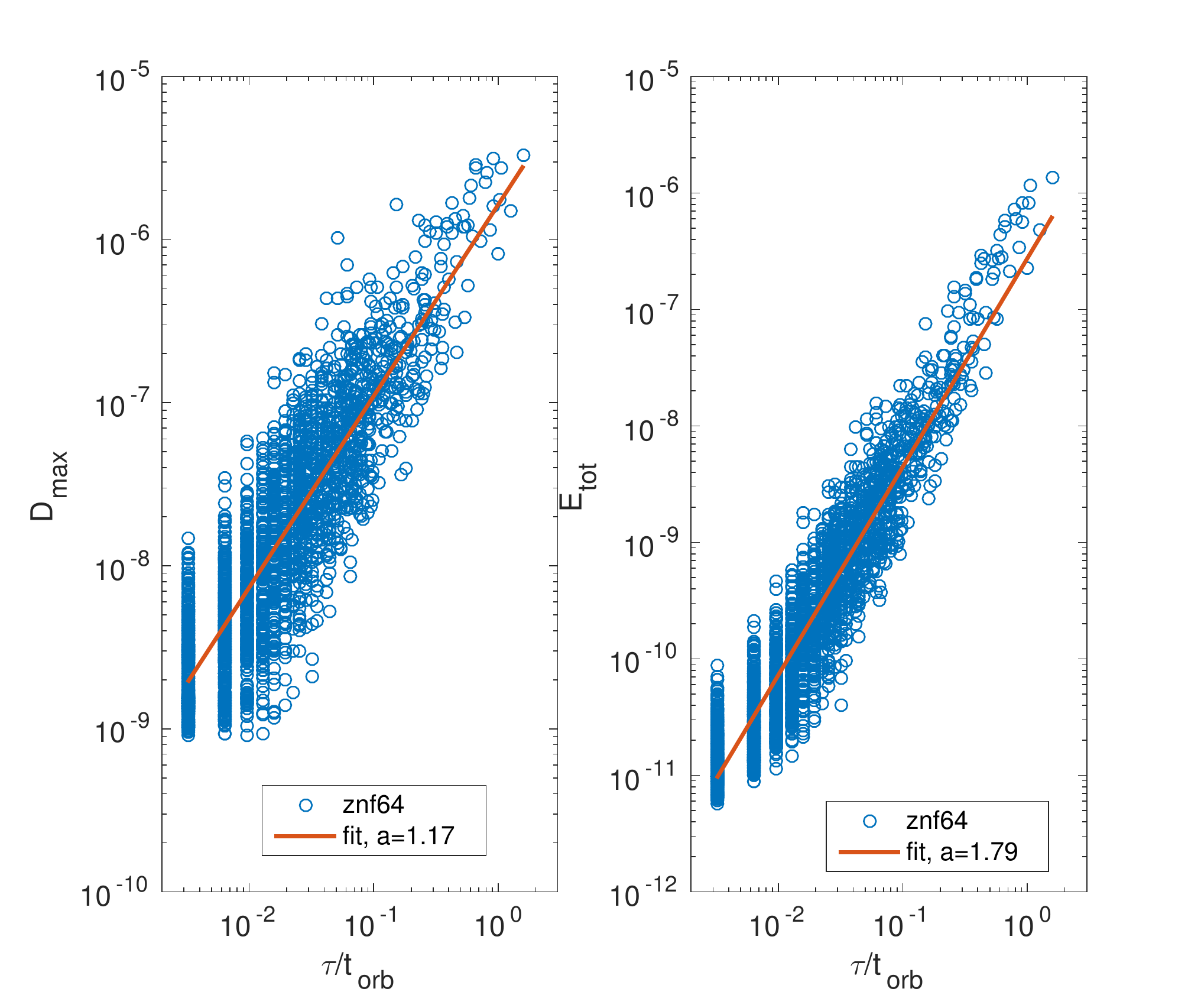}
\caption{Scatter plots showing the correlation between maximum
  dissipation rate $D_\text{max}$ and process lifetime $\tau$ (left panel), and total
  energy dissipated in a process $E_\text{tot}$ and $\tau$ (right
  panel). Superimposed are power law fits with exponent $a$.}
\label{fig::znf64PvT}
\end{figure}

We next check if there is a correlation between
process duration and peak dissipation rate. In the top panel of Figure
\ref{fig::znf64PvT} we see that there is indeed a correlation between
$D_{\text{max}}$ and  $\tau$ which translates across to
 the total power dissipated by a process. 
The persistence of the processes is therefore closely related to the
length and dissipation rate of their component structures. Powerful
dissipative events tend to be associated with spatially larger and
longer-lived processes.
But does the higher energy
dissipation overcome the infrequent occurrence of these long lasting
processes? To explore this we calculate the compensated energy
dissipation rate $E(\tau)\tau$, defined in a similar way to Section
7.3.2. For $\tau/t_\text{orb}>10^{-2}$, $E(\tau)\tau\propto\tau^{1/4}$
indicating that energy is dissipated relatively 
 uniformly over a wide range of process durations. 
Therefore, long lasting, coherent processes
 contribute significantly to the total heating.

\section{Density waves and acoustic radiation}

Up to now we have explored dissipation by current sheets and
turbulent small-scale structures. But what about the compressible
element of the problem, in particular the role of acoustic radiation
and shocks? 
In local MRI turbulence (at least with no net flux), this dissipation route heats the
gas far less effectively, on the whole, but it is nonetheless important to
explore for the following reasons. Pressure appears to be a key
determinant in the MRI's saturation and, consequently, the magnitude
of the turbulent viscous
stress (Ross et al.~2016). But the relationship between
stress and pressure is sensitive to the dissipative particulars of the
gas, at least in local box simulations: 
with explicit diffusion coefficients $\Pi_{xy}\propto P$,
but without $\Pi_{xy}\propto P^{0.5}$. Surprisingly, the
manner in which
diffusion works on the microscales influences the large-scale stress
and its relationship to pressure. How and why this is the case is
an important outstanding problem. The fact that a good fraction of
dissipation occurs in meso-scale, rather than miscroscale, structures
(as shown in Section 3),
must certainly bear on this question and provides one inroad into
thinking about the pressure-stress relationship.

In this section we explore several connections
between sound waves and shocks, on one hand, 
and the dissipative structures, on the other. This work is a first step, and
aims to sketch out certain relationships that future studies might
pursue; it is no way comprehensive. Also, because our primary aim is
explaining the stress-pressure dependency revealed in Ross et
al.~(2016), we use a similar set-up to that paper. Simulations are
diabatic, incorporating viscous and Ohmic heating in addition to a
simple cooling law, and we adopt an ideal gas equation of state (see
Section 2.1). 

\subsection{Small-scale acoustic waves}

\begin{figure}
\centering
\includegraphics[width=8cm]{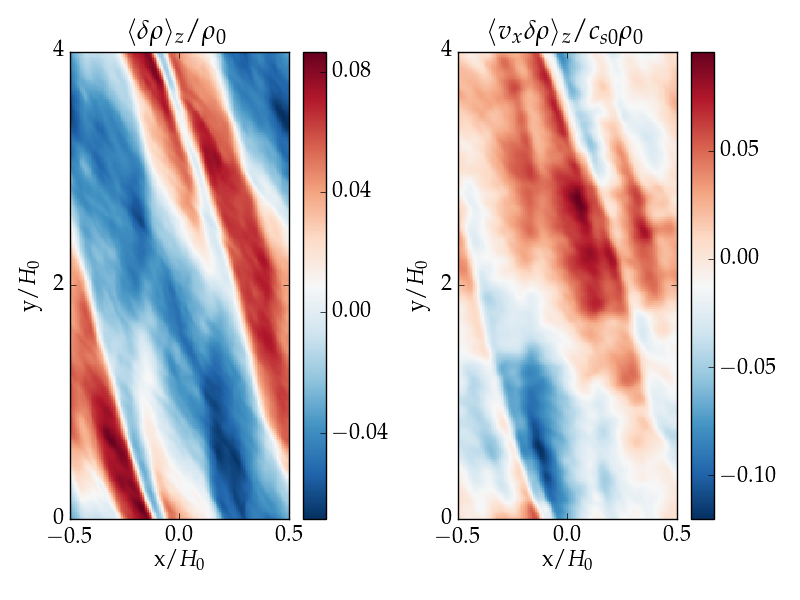}
\caption{The density and radial momentum perturbations, $z$-averaged at
  time $t=35t_\text{orb}$
 showing large scale non-axisymmetric waves. }
\label{Fig::SDW}
\end{figure}

MRI simulations support long-wavelength density waves that continuously
propagate through the box (Heinemann and Papaloizou 2009a, 2009b, 2012); an example is
shown in Figure \ref{Fig::SDW}. But separately to these one can discern
short-scale, short-lived,
and less organised sound waves. We explore these small-scale waves
first.

\begin{figure}
\centering
\includegraphics[width=9cm]{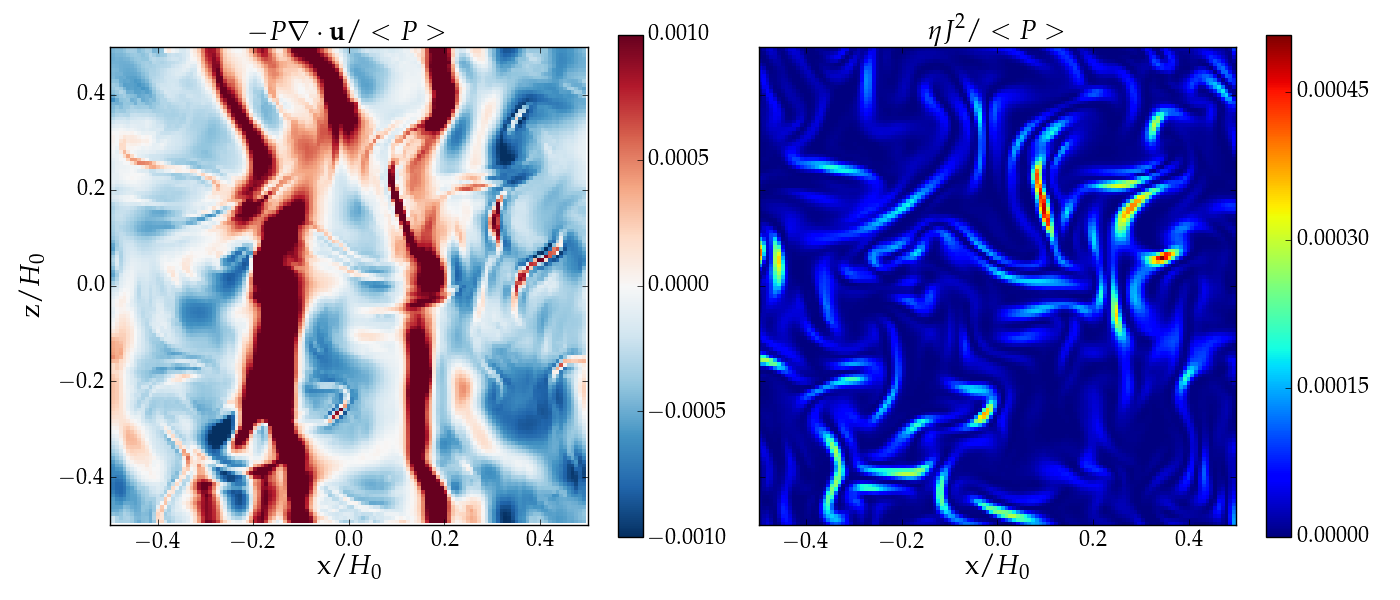}
\caption{$xz$ slices of compression normalised by the box averaged
  pressure (left panel) and $J^{2}$ (right panel)
 at $t=119t_\text{orb}$, the normalised compression work has been saturated at $0.001$.}
\label{Fig::CA2}
\end{figure}

\begin{figure*}
\centering
\includegraphics[width=5.8cm]{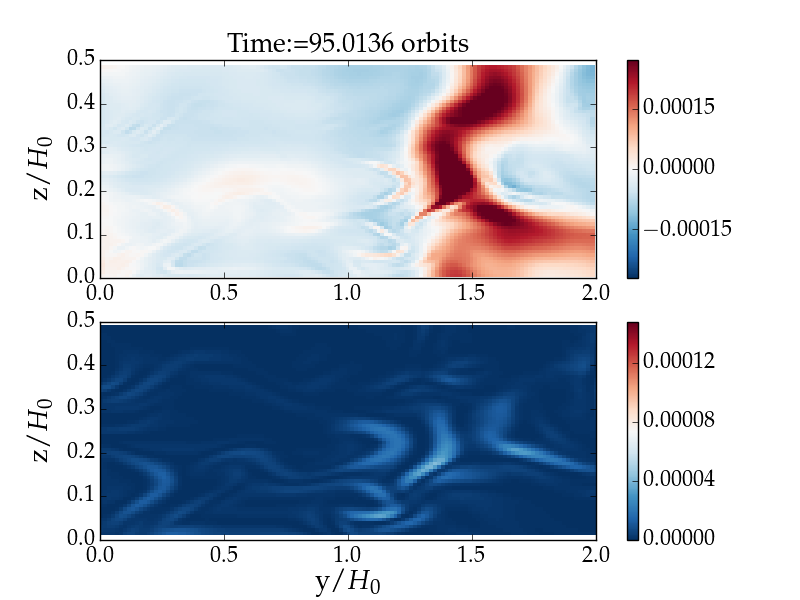}
\includegraphics[width=5.8cm]{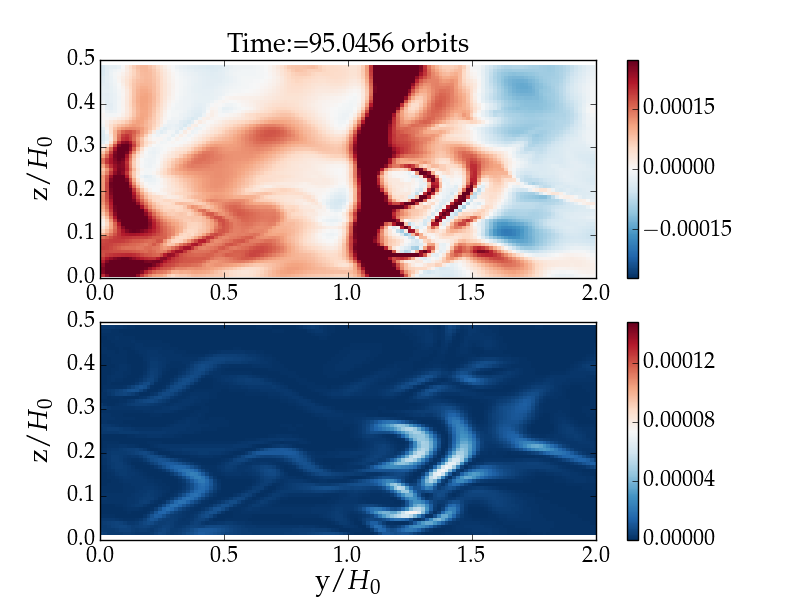}
\includegraphics[width=5.8cm]{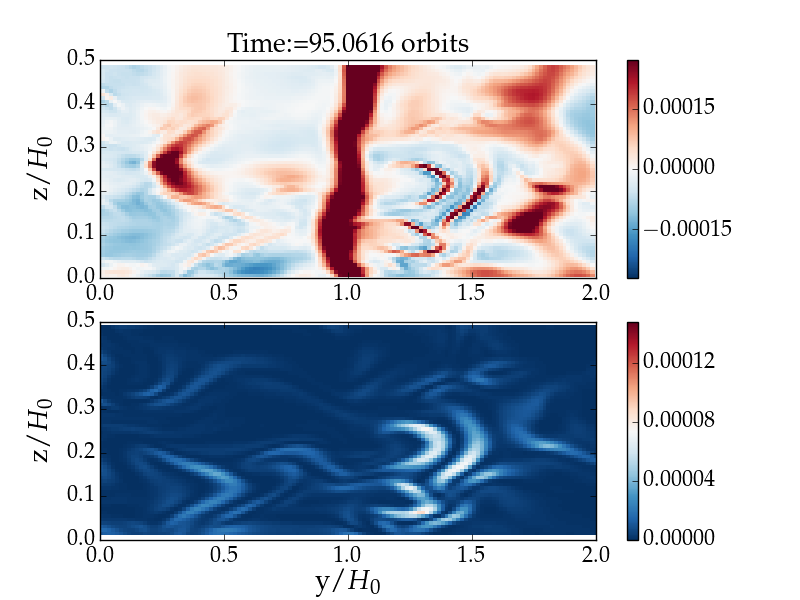}
\caption{Three $yz$-slices of pressure work (top panels) and current
  (bottom panells) 
during the cool phase of the explicit dissipation simulation showing the effect of large-scale shock propagating to the left through dissipative structures (two horse-shoe shaped structures). Both color scales are fixed.}
\label{Fig::yzstructure}
\end{figure*}

 Figure \ref{Fig::CA2} is an
 $xz$-slice
 showing the pressure work $-P\nabla\cdot\boldsymbol{v}$ 
and Ohmic dissipation, both
 normalised by the instantaneous box averaged pressure. The color
 scale for the compression work is saturated at $0.001$ to prevent the
 large-scale density waves dominating the images. These are easy to
 identify as they possess little vertical structure. If we can ignore
 these large waves, one can observe a tangle of
 weaker and shorter-scale filamentary structure in
the pressure work. They correspond to small-scale acoustic waves.
Interestingly, while there is no correlation between $J^{2}$ and the large scale
shocks,
there is a strong correlation
 between the current sheets and the thin
 structures in pressure work. These features and their correlations
are universal, and not limited to this particular snapshot.

It is possible that the dynamics of the current sheets, especially the
reconnection events they undergo, generate the observed
small waves. This would be in line with previous work showing that
accelerations associated with reconnection excite
slow-mode shocks (Priest and
Forbes 1986, Birn and Priest 2007, 
Hillier et al. 2016, Riols and Latter 2016).
On the other hand, the waves might also simply result
from the merging and splitting of paths. Detailed
tracking
of individual current sheets should illuminate how these features
arise.

In the Solar context, at
least, the shocks may provide an effective heating route, though
they appear less important here. It is also unclear how these
small-scale features impact on the larger-scale properties of the
turbulence. Nonetheless they do illustrate one clear connection between MRI
dissipation and the compressible element of the flow. Given that the
longest
and most powerful dissipative structures are meso-scale the importance
of their associated shocks should not be immediately discounted.

\subsection{Large-scale density waves}

Let us now return to the large-scale density waves: given their size they
may be more effective at connecting the large-scale features of the
flow to the small scales. 
If we examine Figure 16, one could argue that the
current sheets distort the density wave as it passes through. 
On the other hand we might expect the density wave to impact on the current
sheet's dynamics. When a shock front
propagates through a region, the fluid is compressed and magnetic
fields clumped by the inflow. Ohmic dissipation and viscous
dissipation will then be enhanced, and may lead to stronger pressure
work in filamentary structures. 

In Figure \ref{Fig::yzstructure} we
show three $yz$ slices separated by $\sim0.01t_\text{orb}$ illustrating the
passage of a density wave through several current sheets. The top panels show
pressure work, while the lower panels show Ohmic dissipation. The
primary density wave, clearly represented by the red vertical `wall' 
in the pressure work,
propagates from right to left. In so doing it passes through three
distinct `boomerang' shaped current sheets located between $y=1$ and
$y=1.5$, and which appear as brighter structures in the Ohmic dissipation.
As can be observed in the second and third snapshots, the dissipation
intensifies as a result of the sheets' interaction with the
wave. Moreover, the interaction results in strong small-scale shocks
enveloping the dissipating boomerangs. Though the detailed nature of
the interaction is difficult to determine, there is clear evidence of
a route by which large-scales interact with small-scale dissipation.

The interaction can be quantified to some extent via the temporal
analysis of Section 4.1. Can we detect the passage of density waves
in the evolution of a dissipative structure's key quantities?
If we consider a process with radial size between
$0.1-0.5H$ we would expect a modulation in the associated pressure
work on a time
scale of $0.1-0.5$ shear times. Indeed, Figure
\ref{fig::znf64Time} shows regular modulations consistent with this. 
Moreover, there is a correlation between the
peaks in the 
dissipation rate (the red curve) and the pressure work (the green curve). This shorter
scale variation is superimposed
on the slower rise and fall of the dissipation. In fact, it is
possible that this longer trend may issue from the cumulative
effect of several density waves.

\section{Conclusion}

Dissipation in MRI turbulence occurs in ribbon-like conglomerates
elongated in the $y$-direction and making an angle $\approx 10^{\circ}$ with the
$y$-axis. The average length of the structures are surprisingly large
($\lesssim H$), with 
the longest extending for multiple $H$ and lasting
for up to approximately one orbit. During this time they appear to be
attacked by an instability
(possibly a Kelvin-Helmholtz or tearing mode), which destroys the conglomerate,
breaking it into smaller pieces that subsequently decay. Meanwhile,
a network of small-scale shock waves is generated, enveloping the
structure. We find clear interactions between dissipative structures and
the large-scale density waves that repeatedly pass through the box.
While the structures distort the waves, the waves intensify
dissipation in the structures, possibly controlling their evolution.

Our spatial results are consistent with ZWBL17,
which considers net-flux simulations in incompressible MRI turbulence at
Reynolds number much greater than those
presented here. The agreement in the main scaling relations
across this range of Reynolds numbers hints that they have reached
the correct asymptotic regime associated with realistic flows. 
In addition, the properties of our simulated current sheets appear to be
closely related to those in reduced MHD (Zhdankin et al.~2016), 
even without the
presence of a net field. Perhaps, the strong toroidal field generated by
the shearing of radial field acts analogously to the guiding
field in RMHD. 

There are several astrophysical consequences that 
follow from the non-negligible sizes of the 
dissipative structures and their
intense energy deposition. 
The dissipation associated with the
longest ribbons may be sufficiently violent and energetic 
to influence observations, cumulatively imparting low-level
variability. 
A separate application is the thermal
processing of chondrules in sufficiently ionised regions of
the Solar nebula. 
The localised and intense heating in these
structures could reach the temperatures necessary to melt dust
coagulations, especially if aided by the `short-circuit' 
instability (Hubbard et al.~2012, McNally et al.~2014).
Further simulations involving realistic cooling are needed
to better estimate the temperature inhomogeneities in the Solar
nebula, and their evolution on short times. In addition, statistics
need to be generated in order to determine how regularly, or irregularly, dust
blobs
encounter dissipating regions of high temperature. Finally, it should
be stated that under normal conditions the MRI is most likely inactive at
the radii where chondrules are thought to be created ($>1$ AU). If
the MRI is responsible for chondrule production it must have been during
short-lived outburst phases in the disk's evolution (Audard et al.~2014).

The size of the dissipating ribbons may bear on the
question of MRI saturation generally which has been shown to depend
on dissipation, via the magnetic Prandtl number or the presence or not
of physical diffusivities. The stress-pressure relation, in
particular, is sensitive to the latter in simulations of zero-magnetic
flux. The simulated interactions between the current sheets and
large-scale density waves might provide one link connecting the
dissipative properties of the flow and its compressible response. In
this paper we merely point out the possibility; future work must
better establish what is going on here and how this connection might
influence the turbulent stress. 
  
Lastly, we would like to make a few caveats regarding our simulations.
Though our scaling laws agree with simulations at greater Re and Rm,
and hence suggest a physical asympototic regime has been achieved,
this is by no means conclusive.
 Current simulations may not yet exhibit a sufficient separation of scales
between $H$ and the shortest dissipation length.
As we push Rm and Re to the very large values associated with hot accretion
flows, it is unclear whether the current sheets retain their elongation in azimuth
(lengths $\lesssim H$), while getting thinner and thinner and narrower
and narrower. If they do, how will their dynamics differ to that described
in this paper? To decide on these questions will be an enormously
expensive but essential exercise. 

\section*{Acknowledgements}

The authors would like to thank the anonymous reviewer for
a positive report that much improved the paper and Pierre Lesaffre who
offered extremely helpful comments on an earlier version of the manuscript. 
This work was partially funded by
STFC grants ST/L000636/1 and ST/K501906/1. Some of the simulations
were run on the DiRAC Complexity system, operated by the
University of Leicester IT Services, which forms part of the STFC
DiRAC HPC Facility (www.dirac.ac.uk). This equipment is funded
by BIS National E- Infrastructure capital grant ST/K000373/1 and
STFC DiRAC Operations grant ST/K0003259/1. DiRAC is part of
the UK National E-Infrastructure.


\begin{thebibliography}{40}

\bibitem{Aud}
Audard, M., Abraham, P., Dunham, M.M., Green, J.D., Grosso, N., 
Hamaguchi, K., Kastner, J.H., Kospal, A., Lodato, G., Romanova, M.M., 
Skinner, S.L., Vorobyov, E.I., Zhu, Z., 2014. In: Beuther, Klessen,
Dullemond,
Henning (eds), Protostars and Planets VI. University of Arizona Press,
Tucson, p387. 


\bibitem{BH91}
Balbus. S. ~A., Hawley, J. ~F., 1991, ApJ, 376, 214.

\bibitem{BH98}
Balbus. S. A., Hawley. J. F., 1998, Rev. Mod. Phy., 70, 1.

\bibitem{B10}
Belloni, T. M., 2010, The Jet Paradigm, Lecture Notes in Physics,
Volume 794, Springer-Verlag Berlin, p53. 

\bibitem{BP07}
Birn, J., Priest, E.R., 2007,
 Reconnection of magnetic fields : magnetohydrodynamics and
 collisionless theory and observations, CUP, Cambridge.

\bibitem{FHT06}
Fromang. S., Hennebelle. P., Teyssier. R., A\&A,  2006, 457, 371-384.

\bibitem{fromang13}
Fromang S., Latter H., Lesur G., \& Ogilvie G. I., 2013, A\&A, 552, A71

\bibitem{FP07}
Fromang. S., Papaloizou. J., A\&A, 2007, 476, 1113.

\bibitem{FPLH07}
Fromang. S., Papaloizou. J., Lesur. G., Heinemann. T., A\&A,  2007, 476, 1123.

\bibitem{G96}
Gammie, C.F., 1996, ApJ, 457, 355.

\bibitem{GL65}
Goldreich. P., Lynden-Bell., MNRAS, 1965, 130, 125


\bibitem{GGSJ09}
Guan. X., Gammie. C. ~F., Simon. J. ~B., Johnson. B. ~M., 2009, ApJ, 694, 1010. 

\bibitem{GO12}
Guilet, J., Ogilvie, G.I., 2012, 424, 2097.
\bibitem{H1}
Heinemann, T. and Papaloizou, J. C. B., 2009a, MNRAS, 397, 52
\bibitem{H2}
Heinemann, T. and Papaloizou, J. C. B., 2009b, MNRAS, 397, 64
\bibitem{H3}
Heinemann, T. and Papaloizou, J. C. B., 2012, MNRAS, 419, 1085


\bibitem{Hill16}
Hillier, A., Takasao, S., Nakamura, N., 2016, AA, 591, 112.
\bibitem{Hub12}
Hubbard, A., McNally, C.P., Mac Low, M., 2012, ApJ, 761, 58.

\bibitem{Hud91}
Hudson, H. S., 1991, Sol. Phys.,
133, 357.

\bibitem{LP12}
Latter. H. N., Papaloizou. J. C. B., 2012, MNRAS, 426, 1107.

\bibitem{LP17}
Latter. H. N., Papaloizou. J. C. B., 2017, MNRAS, 472, 1432.

\bibitem{LL11}
Lesur, G., Longaretti, P.-Y., 2011, AA, 528, 17.
\bibitem{Lour07}
Loureiro, N. F., Schekochihin, A. A., and Cowley, S. C., 2007,
 Physics of Plasmas, 14, 100703


\bibitem{McNally13}
McNally, C. P. Hubbard, A., Mac Low, M, Ebel, D S, D'Alessio, P, 2013.
ApJ, 767, 2. 

\bibitem{MFLJL15}
Meheut. H., Fromang. S., Lesur. G., Joos., Longaretti. P., 2015.,
A\&A, 579, 117.

\bibitem{MK05}
Miyoshi. T., \& Kusano. K., 2005, J. Comput. Phys., 208, 315

\bibitem{NG10}
Nelson R. P., Gressel, O., 2010. MNRAS, 409, 639.

\bibitem{O99}
Ogilvie, G.I., 1999. MNRAS, 304, 557.

\bibitem{PF86}
Priest, E.R., Forbes, T.G., 1986, JGR, 91, 5579.


\bibitem{RmC06}
Remillard, R. A., McClintock, J. E., 2006, ARAA, 44, 49.


\bibitem{RRCLLOH13}
Riols. A., Rincon. F., Cossu. C., Lesur. G., Longaretti. P. ~Y., Ogilvie. G. ~I., Herault. J., Journal of Fluid Mechanics, 2013, 731, 1.

\bibitem{Riols16}
Riols, A., Latter, H., 2016, MNRAS, 460, 2223.

\bibitem{Ross16}
Ross, J., Latter, H.N., Guilet, J., 2016. MNRAS, 455, 526.

\bibitem{Ryan2017}
Ryan, B. R., Gammie, C.F., Fromang, S., Kestener, P., 2017, ApJ, 840, 6.

\bibitem{Sam}
Samtaney, R., Loureiro, N. F., Uzdensky, D. A., Schekochihin, A. A.,
and Cowley, S. C.,
2009, PRL,
103, 105004.

\bibitem{SHB09}
Simon. J.~B., Hawley. J.~F., Beckwith. K., 2009. ApJ, 690, 974.


\bibitem{SG10}
Stone. J., Gardiner. T., ApJSS, 2010, 189, 142.

\bibitem{suresh00}
Suresh. A., 2000, SIAM J. Sci. Comp., 22, 1184


\bibitem{T02}
Teyssier. R., 2002, A\&A, 364, 337.

\bibitem{U10}
Uritsky, V. M., Pouquet, A., Rosenberg, D., Mininni, P. D., and
Donovan, E. F., 2010.
Phys. Rev. E,
82, 056326

\bibitem{Yang}
Yang, L., Zhang, L., He, J., Tu, C., Li, S., Wang, X., Wang, L., 2017,
ApJ, 851, 121. 

\bibitem{YR14}
Yuan, F., Narayan, R., ARAA, 52, 529.

\bibitem{Z13}
Zhdankin, V., Uzdensky, D.A., Perez, J.C., Boldyrev, S., 2013.
ApJ, 771, 124.

\bibitem{Z14}
Zhdankin, V., Boldyrev, S., Perez, J. C., and Tobias, S. M.,
2014. ApJ, 795, 127.
\bibitem{Z15a}
Zhdankin, V., Uzdensky, D. A., and Boldyrev, S., 2015a. PRL, 114,
065002.
\bibitem{Z15b}
Zhdankin, V., Uzdensky, D. A., and Boldyrev, S., 2015b. ApJ, 811, 6.
\bibitem{Z16}
Zhdankin, V., Boldyrev, S., and Uzdensky, D. A., 2016.
 Phys. of Plasmas, 23, 055705.
\bibitem{Z17}
Zhdankin, V., Walker, J., Boldyrev, S., and Lesur, G., 2017, MNRAS,
467, 3620.

\end{thebibliography}
\end{document}